\documentclass[12pt,a4paper]{article}
\usepackage{soul, amssymb,enumitem,booktabs,subfig,todonotes,microtype,setspace,paralist,fullpage, makecell, longtable, bm, xcolor, appendix}
\usepackage[top=1.4in, bottom=1.4in, left=1.35in, right=1.35in]{geometry}
\usepackage{natbib}
\usepackage{url} 
\usepackage[pdftex,colorlinks=true]{hyperref}
\definecolor{darkblue}{rgb}{0,0,.6}
\hypersetup{citecolor=darkblue,linkcolor=darkblue,urlcolor=darkblue}
\usepackage{amsmath}
\usepackage{amsfonts}
\usepackage{epsfig,csquotes}
\usepackage{graphics}
\usepackage{palatino,eulervm}
\usepackage{graphicx}
\usepackage{lscape,dsfont}
\usepackage{rotating}
\usepackage{orcidlink}
\usepackage[linewidth=1pt]{mdframed}
\usepackage{hyperref}
\allowdisplaybreaks[4]

\newcommand{\x}{\mathcal{X}}

\providecommand{\U}[1]{\protect\rule{.1in}{.1in}}
\renewcommand{\baselinestretch}{1.1}
\graphicspath{{plots/}}

\newcommand{\Rlogo}{\protect\includegraphics[height=1.8ex,keepaspectratio]{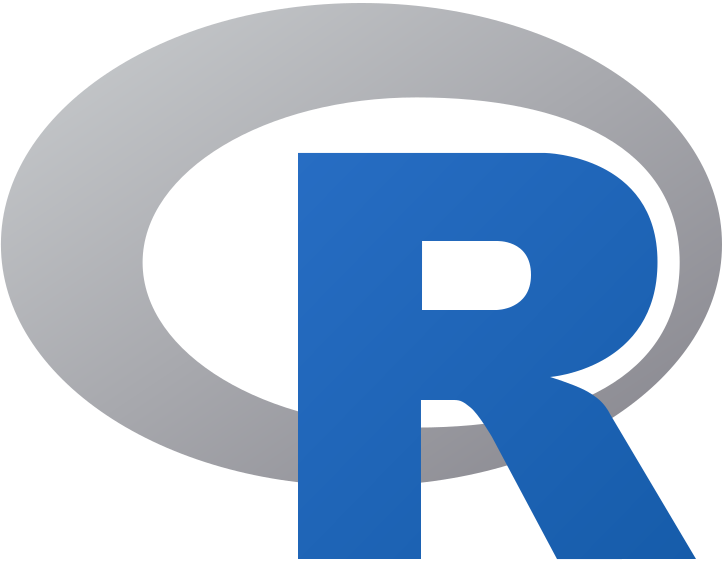}}

\usepackage{amsthm,thmtools}
\declaretheorem{theorem}

\makeatletter
\def\th@newremark{\th@remark\thm@headfont{\bfseries}}
\makeatletter
\theoremstyle{newremark}

\declaretheoremstyle[
  spaceabove=6pt, spacebelow=6pt,
  headfont=\bfseries,
  notefont=\mdseries, notebraces={(}{)},
bodyfont=\normalfont,
  postheadspace=0.5em,
]{mystyle}

\begin{document}

\def\spacingset#1{\renewcommand{\baselinestretch}{#1}\small\normalsize} \spacingset{1}

\title{\Large \bf A nonlinearity and model specification test for functional time series}
\author{
Xin Huang\orcidlink{0000-0003-0176-1197} \qquad Han Lin Shang\orcidlink{0000-0003-1769-6430} \footnote{Corresponding address: Department of Actuarial Studies and Business Analytics, Macquarie University, Sydney, NSW 2109, Australia; Telephone: +61(2) 9850 4689; Email: hanlin.shang@mq.edu.au} \qquad Tak Kuen Siu\orcidlink{0000-0003-2823-5138}
\\
    Department of Actuarial Studies and Business Analytics \\
    Macquarie University
}

\date{}

\maketitle

\begin{abstract}

An important issue in functional time series analysis is whether an observed series comes from a purely random process. We extend the BDS test, a widely-used nonlinear independence test, to the functional time series. Like the BDS test in the univariate case, the functional BDS test can act as the model specification test to evaluate the adequacy of various prediction models and as a nonlinearity test to detect the existence of nonlinear structures in a functional time series after removing the linear structure exhibited. We show that the test statistic from the functional BDS test has the same asymptotic properties as those in the univariate case and provides the recommended range of its hyperparameters. Additionally, empirical data analysis features its applications in evaluating the adequacy of the fAR(1) and fGARCH(1,1) models in fitting the daily curves of cumulative intraday returns (CIDR) of the VIX index. We showed that the functional BDS test remedies the weakness of the existing independence test in the literature, as the latter is restricted in detecting linear structures, thus, can neglect nonlinear temporal structures.
\vspace{.1in}

\end{abstract}

\medskip

\spacingset{1.54}

\newpage

\section{Introduction}

Functional time series analysis is a fusion between functional data and time series analyses. Similar to univariate and multivariate time series, a temporal dependence structure exists in functional observations, manifesting themselves in a graphical form of curves, images, or shapes. Functional time series can typically be classified into two categories: First is segmenting a univariate time series into (sliced) functional time series. For instance, \cite{RWZ21} consider intraday volatility to form functional time series $\Big[\x_{1}(u),$$\x_{2}(u)$$,\ldots,$$\x_{N}(u)\Big]$ defined for a continuum $u \in [u_{1},u_{p}]$. The other category is when the continuum is not a time variable, such as age \citep[see, e.g.,][]{SHX22} or wavelength in spectroscopy \citep[see, e.g.,][]{SCS22}.

While functional time series are continuous objects, discrete time series are scalar-valued. Functional time series enjoy at least three advantages:
\begin{inparaenum}
\item[1)] Compared to univariate time series, functional representation of time-series data accommodates well with data collected at ultra-high-frequency and can alleviate the burden of the parameter estimation that commonly arises in modeling a large volume of observations. 
\item[2)] As by-products, functional derivatives can provide additional insight into the data under investigation \citep[see, e.g.,][]{shang2019visualizing, Hooker2020}. 
\item[3)] Some data are more natural to be considered functions that behave like smooth curves rather than separated discrete observations, such as age-specific mortality data.
\end{inparaenum}

Over the past two decades, there have been rapid developments in functional time series analysis. \cite{ramsay2002applied, ramsey2005functional} provide comprehensive overviews of the major advances and the fundamental concepts and techniques used in functional data analysis. An important branch of developments of functional time series analysis is extending the mainstream models and analytical tools in univariate time series to functional cases \citep[see, e.g.,][]{kokoszka2017introduction}. To name a few, \cite{kokoszka2017inference} and \cite{mestre2021functional} proposed a functional autocorrelation function (fACF) to quantify linear serial correlation in a functional time series. \cite{HS23} proposed a nonlinear fACF to measure nonlinear dependence in a functional time series.

\cite{bosq1991modelization} extended the autoregressive (AR) model to the functional case, referred to as the fAR model. Since then, many functional time series models are extended from the fAR model. They include the autoregressive Hilbertian model with exogenous variables (ARHX) model \citep{damon2002inclusion}, the Hilbert moving average model \citep{turbillon2007estimation}, the functional autoregressive moving average (fARMA) model \citep{klepsch2017prediction}, the seasonal functional autoregressive model \citep{zamani2022seasonal}, and the seasonal autoregressive moving average Hilbertian model with exogenous variables (SARMAHX) model \citep{gonzalez2017forecasting}. For modeling conditional variance, they include the functional autoregressive conditional heteroskedasticity (fARCH) model \citep{hormann2013functional}, functional generalized autoregressive conditional heteroskedasticity (fGARCH) model \citep{aue2017functional}, and fGARCH-X model \citep{RWZ21}.

Despite the increasing interest and research on functional time series, many existing measures/tools that study the structure underlying the observed functional time series, including the fACF, are based on the autocovariance and/or autocorrelation \citep{horvath2013test,zhang2016white,mestre2021functional}. They only capture the linear temporal structure. Except for nonlinear fACF in \cite{HS23}, there is little research to study the nonlinear temporal structures within the functional time series literature. Additionally, since linear structures restrict those tools, they cannot test all possible deviations from randomness. Therefore, a robust model specification test must be improved to evaluate the adequacy of functional time series models. 

In this paper, we extend the BDS test to functional time series. Just like the univariate case, the proposed test can be used as an IID test on estimated residuals to evaluate the adequacy of the fitted model and as a nonlinearity test on residuals of functional time series after removing linear temporal structures exhibited in the investigated data.

The BDS test proposed by \cite{broock1996test} is the most widely used nonlinearity test and model specification test in univariate time series analysis. The reason behind its popularity is many folds: First, the BDS test requires minimal assumptions and previous knowledge about the investigated data sets. When the BDS test is applied to model residuals, the asymptotic distribution of its test statistic is independent of estimation errors under certain sufficient conditions. Specifically, \cite{f1996nuisance} shows that for linear additive models or models that can be transformed into that form, the BDS test is nuisance parameter-free and does not require any adjustment when applied to fitted model residuals. 

Second, the BDS test tests against various forms of deviation from randomness. While the null hypothesis of the BDS test is the investigated time series is generated by an IID process, its alternative hypothesis is not specified. It may be thought of as a portmanteau test. This implies that the BDS test can detect any non-randomness exhibited in the investigated time series. Additionally, a fast algorithm exists for computing BDS test statistics, which ensures the BDS test's easy and speedy application on empirical applications \citep{lebaron1997fast}. Also, the BDS statistic asymptotic distribution theory does not require higher-order moments to exist. This property is especially useful in financial time series analysis since many financial time series exhibit heavy-tailed distributions whose higher-order moments may not exist.

The rest of this paper is structured as follows. Section~\ref{section2} provides the specification of the functional BDS test. In appendix \ref{appendix2}, we provide detailed proof of the asymptotic distribution of the test statistics of the functional BDS test. In Section~\ref{section4}, we present Monte-Carlo experiments on the IID functional time series and simulated fAR(1) functional time series to provide the recommended dimension and distance hyperparameters range. In Section~\ref{section5}, the functional BDS test is used to test the adequacy of the fAR(1) model and fGARCH(1,1) model of daily curves of intraday VIX index returns. Conclusions are given in Section~\ref{section6}, along with some ideas on how the methodology presented here can be further extended.  

\section{BDS Test for functional time series}\label{section2}

The BDS test uses ``correlation integral", a popular measure in chaotic time series analysis. According to \cite{packard1980geometry} and \cite{takens1981detecting}, the method of delays can embed a scalar time series $\{x_{i}:i=1,2,...,N\}$ into a $m$-dimensional space as follows
\begin{equation*}
\vec{x}_{i}=(x_{i},x_{i+1},...,x_{i+m-1}), \qquad \vec{x}_{i} \in R^{m}.
\end{equation*}
Accordingly, $\vec{x}_{i}$ is called $m$-history of $x_{i}$. \cite{grassberger1983characterization} proposed correlation integral as a measure of the fractal dimension of deterministic data as it records the frequency with which temporal patterns are repeated. The correlation integral at the embedding dimension $m$ is given by
\begin{align}\label{eq:BDS1}
C(m,N,r) &=\frac{2}{M(M-1)}\sum_{1\leq i<j\leq M} \Theta(r-\Vert \vec{x}_{i}-\vec{x}_{j} \Vert ), \quad r>0
\\ 
\Theta(a) &=0, \quad  \text{if} \; a \leq 0 \nonumber \\
\Theta(a) &=1, \quad  \text{if} \; a > 0 \nonumber
\end{align}
where $N$ is the size of the data sets, $M=N-m+1$ is the number of embedded points in m-dimensional space, $r$ is the distance used for testing the proximity of the data points, and $\Vert \cdot \Vert$ denotes the sup-norm. In essence, $C(m, N, r)$ measures the fraction of the pairs of points $\vec{x}_{i}$, $i=1, 2, \ldots, M$, the sup-norm separation of which is less than $r$. 


\cite{brock1987notes} showed that under the null hypothesis that $\{x_{i}:i=1,2,...,N\}$ are IID with a non-degenerated distribution function $F()$,
\begin{align*}
C(m, r) &:= \lim_{N \to \infty} C(m, N, r) \\
C(m, r)&\rightarrow C^{m}(1,r) \quad \text{with probability 1.} 
\end{align*}
According to \cite{broock1996test}, the BDS statistic for $m>1$ is defined as
\begin{align*} \label{eq:BDS}
\text{BDS}(m, M, r)=\frac{\sqrt{N}}{\sigma}\left[C(m, N, r)-C^{m}(1,r)\right]
\end{align*}where $M=N-m+1$, 
\begin{equation*}
\sigma^{2}=4\left(K^{m}+2\sum^{m-1}_{j=1}K^{m-j}C^{2j}+(m-1)^{2}C^{2m}-m^{2}KC^{2m-2}\right),
\end{equation*}
$C=\int[F(z+r)-F(z-r)]dF(z)$ and $K=\int[F(z+r)-F(z-r)]^{2}dF(z)$. 
Note that $C(1,N,r)$ is a consistent estimate of $C$, and $K$ can be consistently estimated by
\begin{equation}\label{eq:BDS2}
\frac{6}{M(M-1)(M-2)}\sum_{1<t<s<u<M}\Theta(r-\Vert \vec{x}_{t}-\vec{x}_{s} \Vert )\Theta(r-\Vert \vec{x}_{s}-\vec{x}_{u} \Vert ).
\end{equation}
Under the IID hypothesis, $\text{BDS}(m, M, r)$ has a limiting standard normal distribution as $M\rightarrow \infty$. 


The above specification of the BDS test is for scalar time series. When the object is functional time series, one needs to adjust the computation of sup-norm separation of the $m$-histories in~\eqref{eq:BDS1} and~\eqref{eq:BDS2}.

Given a functional time series $\bm{\x}(u)=\{\x_{t}(u);u\in [u_{1},u_{p}],t=1,2,\ldots,N\}$, the $m$-history of $\x_{i}(u)$ is constructed by its $m$ neighbouring observations, namely
\begin{equation*}
\overrightarrow{\x_{i}(u)}=\left[\x_{i}(u),\x_{i+1}(u),...,\x_{i+m-1}(u)\right].
\end{equation*}
The sup-norm of two sets of $m$ functions can be measured by taking the maximum distance between the corresponding curves. Specifically, if we use $L_2$ norm as the distance measure between two curves, 
\begin{equation*}
\Vert \overrightarrow{\x_{i}(u)}-\overrightarrow{\x_{j}(u)} \Vert=\max(\Vert \x_{i}(u)-\x_{j}(u) \Vert_{2},\ldots, \Vert \x_{i+m-1}(u)-\x_{j+m-1}(u) \Vert_{2}).
\end{equation*}

Since we adjust the specification of the BDS test statistic to be adaptive to the functional case, to determine the critical value of the BDS test after the adjustment, one needs to derive its asymptotic distribution under the null hypothesis. In appendix~\ref{appendix2}, we prove that the asymptotic normality for the univariate BSD test statistic is also valid for the functional case. Indeed, the asymptotic normality result presented in Theorem~1 of Appendix~\ref{appendix2} is quite versatile. It holds for any norm $||\cdot||_{\cal H}$ on a separable Hilbert space ${\cal H}$, which is more general than the $L^2$-norm. 

The $L_2$ norm is not the only distance measure of two functions. Other common choices include $L_1$ norm and $L_{\text{inf}}$ norm. All of them, including other norms, can be used for computing the sup-norm of $m$-histories of functional time series. However, the choice of the distance measure determines the recommended range of distance hyperparameter $r$ as well as the speed of convergence of the test statistic and the power of the test. In Section~\ref{section4}, we present power and size experiments on random and structured functional time series when $L_1$, $L_2$ and $L_{\text{inf}}$ are selected as the distance measure inside the sup-norms.

\section{Monte-Carlo experiments}\label{section4}

In the section, we conduct Monte-Carlo experiments on simulated IID and structured functional time series to provide the recommended range of hyperparameters of the functional BDS test, namely $m$, $r$, and the preferred norms inside the sup-norms.

We use two metrics to evaluate the selection of the hyperparameters and the norms: 
\begin{inparaenum}
\item[1)] the resemblance of normality of the test statistics on the IID process; and
\item[2)] the power of rejecting $H_{0}$ on the structured process. 
\end{inparaenum}

For the resemblance of normality, we simulated 200 paths of 500 IID functional time series and computed the BDS test statistic on each path with $m=(2,3,\ldots,10)$ and $r=(0.25\text{s.d.},0.5\text{s.d.} \ldots,2\text{s.d.})$. Table~\ref{tb:BDS2} provides the $p$-value of the Kolmogorov–Smirnov (KS) test for each combination of $m$ and $r$ when $L_2$ is selected as the norm inside the sup-norms. The respective tables with $L_1$ and $L_{\text{inf}}$ being selected as the norms inside the sup-norms are provided in the appendix. The KS test examines against the null hypothesis that the computed functional BDS test statistic is from a standard normal distribution. A $p$-value less than 0.025 (highlighted in red) indicates the rejection of $H_{0}$, which means the generated BDS test statistics cannot be assumed to follow a standard normal distribution. On the contrary, the higher the $p$-value is, the closer the BDS test statistics are to a standard normal distribution.  

\renewcommand{\arraystretch}{1.1}
\begin{table}[!htbp]
\tabcolsep 0.22in
\centering
\caption{\small{The $p$-value of the KS test on functional BDS test statistics with $L_{2}$ norm computed on 200 paths of 500 simulated IID functional time series.}}
\begin{tabular}{@{}lrrrrrrrrr@{}}
\toprule  
	& \multicolumn{9}{c}{$m$} \\
 $r$ &  $2$ &  $3$ &  $4$ &  $5$ &  $6$ &  $7$ & 
 $8$ &  $9$ &  $10$\\ \midrule
$0.25\text{s.d.}$ &
  {\color[HTML]{FE0000} 0.00} &
  {\color[HTML]{FE0000} 0.00} &
  {\color[HTML]{FE0000} 0.00} &
  {\color[HTML]{FE0000} 0.00} &
  {\color[HTML]{FE0000} 0.00} &
  {\color[HTML]{FE0000} 0.00} &
  {\color[HTML]{FE0000} 0.00} &
  {\color[HTML]{FE0000} 0.00} &
  {\color[HTML]{FE0000} 0.00}   \\
$0.5\text{s.d.}$ &
  {\color[HTML]{FE0000} 0.01} &
  {\color[HTML]{FE0000} 0.00} &
  {\color[HTML]{FE0000} 0.00} &
  {\color[HTML]{FE0000} 0.00} &
  {\color[HTML]{FE0000} 0.00} &
  {\color[HTML]{FE0000} 0.00} &
  {\color[HTML]{FE0000} 0.00} &
  {\color[HTML]{FE0000} 0.00} &
  {\color[HTML]{FE0000} 0.00}       \\
$0.75\text{s.d.}$  &
  {\color[HTML]{FE0000} 0.00} &
  0.03 &
  0.06 &
  {\color[HTML]{FE0000} 0.00} &
  {\color[HTML]{FE0000} 0.00} &
  {\color[HTML]{FE0000} 0.00} &
  {\color[HTML]{FE0000} 0.00} &
  {\color[HTML]{FE0000} 0.00} &
  {\color[HTML]{FE0000} 0.00}      \\
$\text{s.d.}$  &
  0.84 &
  0.63 &
  0.06 &
  {\color[HTML]{FE0000} 0.00} &
  0.26 &
  {\color[HTML]{FE0000} 0.01} &
  {\color[HTML]{FE0000} 0.00} &
  {\color[HTML]{FE0000} 0.00} &
  {\color[HTML]{FE0000} 0.00}      \\
$1.25\text{s.d.}$   &
  0.69 &
  0.52 &
  0.19 &
  0.88 &
  0.10 &
  0.03 &
  {\color[HTML]{FE0000} 0.01} &
  {\color[HTML]{FE0000} 0.00} &
  {\color[HTML]{FE0000} 0.02}      \\
$1.5\text{s.d.}$  &
  0.56 &
  0.44 &
  0.38 &
  0.03 &
  {\color[HTML]{FE0000} 0.00} &
  0.23 &
  0.60 &
  0.03 &
  0.94       \\
$1.75\text{s.d.}$ &
  0.34 &
  0.37 &
  0.33 &
  0.14 &
  0.19 &
  0.22 &
  0.07 &
  0.06 &
  0.06       \\
$2\text{s.d.}$ &
  0.27 &
  0.00 &
  0.29 &
  0.93 &
  {\color[HTML]{FE0000} 0.01} &
  0.33 &
  0.52 &
  {\color[HTML]{FE0000} 0.02} &
  {\color[HTML]{FE0000} 0.02}      \\ \bottomrule
\end{tabular}
\label{tb:BDS2}
\end{table}

From the result in Tables~\ref{tb:BDS2} and~\ref{tb:BDS1}, we can see that the functional BDS test with a moderate $m$ ($2 \leq m \leq7$) and a sufficiently large $r$ ($r>=\text{s.d.}$) ensures that the respective test statistics have distributions sufficiently close to a standard normal distribution under the null hypothesis.   

For the power test, we simulated 200 paths of an fAR(1) process. In each path, we generate 500 observations, and each functional observation is formed by 100 equal-spaced points within (0,1). Following the definition given in \cite{bosq2000linear}, let $\mathcal{H}$ be a separable Hilbert space with norm $\| \cdot \|$ and scalar product $\langle\cdot,\cdot\rangle$, a sequence $\{\x_{1}(u),\ldots,\x_{N}(u)\}$ of $\mathcal{H}$ random variables is called an fAR(1) associated with $(\mu,\epsilon,\rho)$ if
\begin{equation}\label{FAR1}
\x_{t+1}(u)-\mu(u)=\rho[\x_{t}(u)-\mu(u)]+\epsilon_{t}(u),
\end{equation}
where the linear operator $\rho$ ($||\rho||<1$) acting on the Hilbert space ${\cal H}$ is compact, and a set of random error terms $\left\{\epsilon_{1}(u), \epsilon_{2}(u), \ldots,\epsilon_{N}(u)\right\}$ is a set of independent random variables with a distribution satisfying the conditions: $E\left[\epsilon_{t}(u)\right]=0$, $0<E\left[\epsilon^{2}_{t}(u)\right]<\infty$, and $E\left[\epsilon_{t}(u)\epsilon_{s}(u)\right]=\sigma^{2}\delta_{t,s}$, for $t,s\in \mathbb{Z}$ ($\delta_{t,s}=1$ if $t=s$; $\delta_{t,s}=0$ if $t\neq s$). Equation~\eqref{FAR1} can be easily extended to define a fAR($p$) process by including additional terms of the form $\rho_{k}[\x_{t-k}(u)-\mu(u)]$. We use the \verb|simul.far| function in the \enquote*{far} package \citep{serge22} in \Rlogo \ \citep{Team22} to generate the fAR(1) process. The simulated fAR(1) process has the linear operator $\rho=0.1$. The error terms $\epsilon_{t}(u)$ are strong white noise, and five equally-spaced sinusoidal bases \{1, $\sqrt{2}\sin (2\pi u)$, $\sqrt{2}\cos (2\pi u)$, $\sqrt{2} \sin (4\pi u)$, $\sqrt{2} \cos (4\pi u)$\} are used to construct the fAR(1) model. We specifically choose a small $\rho$ so the generated process has relatively weak temporal structures.

After simulating 200 paths of the fAR(1) process, we compute the functional BDS test statistic on each simulated functional time series. Table~\ref{tb:BDSN2} presents the probability that the functional BDS test successfully rejects the IID hypothesis on a structured process when $L_2$ is selected as the norm. Table~\ref{tb:BDSN1} in the appendix provides the respective tables with $L_1$ and $L_{\text{inf}}$. A value of 100\% indicates that the BDS test made correct inferences at all simulated paths, whereas a value less than 100\% suggests it failed to distinguish a structured process from a random one at certain paths. The value less than 100\% is highlighted in red, indicating that this particular combination of $m$ and $r$ impairs the power of the test.

\begin{table}[!htbp]
\tabcolsep 0.17in
\centering
\caption{\small{The successful rejection rate of the functional BDS test on 200 paths of the simulated fAR(1) process ($\rho=0.1$) of 500 observations with $L_{2}$ being used as the norm inside the sup-norms and different choices of $m$ and $r$.}}
\begin{tabular}{@{}lrrrrrrrrr@{}}
\toprule  
	& \multicolumn{9}{c}{$m$} \\
 $r$ &  $2$ &  $3$ &  $4$ &  $5$ &  $6$ &  $7$ & 
 $8$ &  $9$ &  $10$\\ \midrule
$0.25\text{s.d.}$  & 100\% &
  {\color[HTML]{FE0000} 99\%} &
  100\% &
  100\% &
  {\color[HTML]{FE0000} 98\%} &
  {\color[HTML]{FE0000} 68\%} &
  {\color[HTML]{FE0000} 20\%} &
  {\color[HTML]{FE0000} 3\%} &
  {\color[HTML]{FE0000} 0\%}   \\
$0.5\text{s.d.}$   &
  100\% &
  100\% &
  100\% &
  100\% &
  100\% &
  {\color[HTML]{FE0000} 99\%} &
  {\color[HTML]{FE0000} 93\%} &
  {\color[HTML]{FE0000} 52\%} &
  {\color[HTML]{FE0000} 15\%}       \\
$0.75\text{s.d.}$  &
  100\% &
  100\% &
  100\% &
  100\% &
  100\% &
  100\% &
  100\% &
  {\color[HTML]{FE0000} 99\%} &
  {\color[HTML]{FE0000} 99\%}      \\
$\text{s.d.}$  &
  100\% &
  100\% &
  100\% &
  100\% &
  100\% &
  100\% &
  100\% &
  100\% &
  100\%       \\
$1.25\text{s.d.}$   &
  100\% &
  100\% &
  100\% &
  100\% &
  100\% &
  100\% &
  100\% &
  100\% &
  100\%      \\
$1.5\text{s.d.}$   &
  100\% &
  100\% &
  100\% &
  100\% &
  100\% &
  100\% &
  {\color[HTML]{FE0000} 99\%} &
  {\color[HTML]{FE0000} 99\%} &
  {\color[HTML]{FE0000} 98\%}      \\
$1.75\text{s.d.}$ &
  {\color[HTML]{FE0000} 98\%} &
  100\% &
  {\color[HTML]{FE0000} 99\%} &
  {\color[HTML]{FE0000} 98\%} &
  {\color[HTML]{FE0000} 98\%} &
  {\color[HTML]{FE0000} 98\%} &
  {\color[HTML]{FE0000} 99\%} &
  {\color[HTML]{FE0000} 97\%} &
  {\color[HTML]{FE0000} 96\%}     \\
$2\text{s.d.}$ &
  {\color[HTML]{FE0000} 98\%} &
  {\color[HTML]{FE0000} 95\%} &
  {\color[HTML]{FE0000} 96\%} &
  {\color[HTML]{FE0000} 92\%} &
  {\color[HTML]{FE0000} 90\%} &
  {\color[HTML]{FE0000} 89\%} &
  {\color[HTML]{FE0000} 92\%} &
  {\color[HTML]{FE0000} 93\%} &
  {\color[HTML]{FE0000} 86\%}      \\ 
  \bottomrule
\end{tabular}
\label{tb:BDSN2}
\end{table}

In Tables~\ref{tb:BDSN2} and~\ref{tb:BDSN1}, the results of the power experiments indicate the BDS test attains the highest successful rejection rate when $m$ is between 2 and 7 and $r$ is between $0.5\text{s.d}$ and $1.5\text{s.d.}$. 

For the robustness experiments, we randomly replaced  1\% of the simulated IID functional time series to have a distinctive higher mean than the rest of the observations and then repeated the normality resemblance experiments. Table~\ref{tb:sensitivity} that presents the $p$-values of the KS test for $L_{1}$, $L_{2}$, and $L_{\text{inf}}$ is given in the appendix. The results showed that including random outliers does not impair the convergence to normality for the functional BDS test when $L_{1}$ and $L_{2}$ are used as the norm inside the sup-norms. However, when $L_{\text{inf}}$ is used inside the sup-norms, the test is significantly affected by the outliers. The presence of outliers makes the generated test statistic fail the KS test for most of the combinations of $m$ and $r$ when $L_{\text{inf}}$ is chosen as the norm inside the sup-norm. 

Lastly, we performed a size experiment to guide the preferred length of the functional time series so that the BDS test has satisfactory performance. We repeat the normality resemblance experiment and the power test with $m=3$ and $r=\text{s.d.}$ where the selected $m$ and $r$ are within the recommended range as indicated by our previous normality resemblance experiments and power test. The simulated functional time series length is 100, 250, 500, 750, or 1000. The result of the normality resemblance experiment is presented in Table~\ref{tb:BDSM1}, and the power test result is given in Table~\ref{tb:BDSM2}. The size experiment indicates that with an appropriate selection of $m$ and $r$, the functional BDS test has satisfactory performance for functional time series with a length greater or equal to 250.

\begin{table}[!htbp]
\tabcolsep 0.24in
\centering
\caption{\small{The KS test $p$-value of the functional BDS test statistic on IID functional time series with various lengths.}}
\begin{tabular}{@{}lrrrrr@{}}
\toprule
KS test $p$-value ($m=3$, $r=\text{s.d.}$) 	& n=100 & n=250 & n=500 & n=750 & n=1000 \\
\midrule
$L_{1}$                              				& {\color[HTML]{FE0000} 0.00} & 0.10                        & 0.13                        & 0.01  & 0.43   \\
$L_{2}$                              				& {\color[HTML]{FE0000} 0.00} & 0.50                        & 0.43                        & 0.81  & 0.81   \\
$L_{\text{inf}}$                   				& 0.06                    & {\color[HTML]{FE0000} 0.00} & {\color[HTML]{FE0000} 0.01} & 0.43  & 0.13 \\ 
\bottomrule
\end{tabular}
\label{tb:BDSM1}
\end{table} 

\begin{table}[!htbp]
\tabcolsep 0.25in
\centering
\caption{\small{The successful rejection rate of the functional BDS test on simulated fAR(1) process ($\rho=1$) with various lengths.}}
\begin{tabular}{@{}lrrrrr@{}}
\toprule
Rejection rate ($m=3$, $r=\text{s.d.}$) 	& n=100 & n=250 & n=500 & n=750 & n=1000 \\
\midrule
$L_{1}$                      				& {\color[HTML]{FE0000} 88\%} & 100\% & 100\% & 100\% & 100\%  \\
$L_{2}$                      				& {\color[HTML]{FE0000} 88\%} & 100\% & 100\% & 100\% & 100\%  \\
$L_{\text{inf}}$           				& {\color[HTML]{FE0000} 94\%} & 100\% & 100\% & 100\% & 100\%  \\ 
\bottomrule
\end{tabular}
\label{tb:BDSM2}
\end{table}

To conclude, to ensure the convergence of normality and the power of the test, we recommend the dimension hyperparameter $m$ to be in the range of 2 and 7, and the distance parameter $r$ between $\text{s.d}$ and $1.5\text{s.d.}$. For the norm inside the sup-norms, we recommend $L_{1}$ and $L_{2}$ as they are more robust to outliers. Lastly, a functional time series with more than 250 observations is recommended for the functional BDS test to perform satisfactorily.

\section{Evaluation of the adequacy of the fAR(1) model and fGARCH(1,1) model on VIX tick returns}\label{section5}

This section depicts an empirical application of the functional BDS test to evaluate the adequacy of the fAR(1) model and the fGARCH(1,1) model in fitting the daily curves of intraday VIX (volatility) index returns. VIX is a forward-looking volatility measure of the future equity market based on a weighted portfolio of 30-day S\&P 500 Index option prices. The VIX index is a key measure of risk for the market. Therefore, accurately predicting the VIX index is essential in risk management, especially for hedge and pension funds.

Most existing studies that attempted to model and predict the VIX index treat it as a discrete time series. To name a few, \cite{KST08} used an autoregressive fractionally integrated moving average (ARFIMA) model, and \cite{FMS14} employed a heterogeneous AR model to predict future VIX index. Recently, the functional time series model has provided new alternatives to extract additional information underlying the VIX dynamics and potentially provides more accurate forecasts for the market expectation for the equity risk in the future \citep[see, e.g.,][]{SK19}.

\begin{figure}[!htb]
\centering
\includegraphics[width=13.5cm]{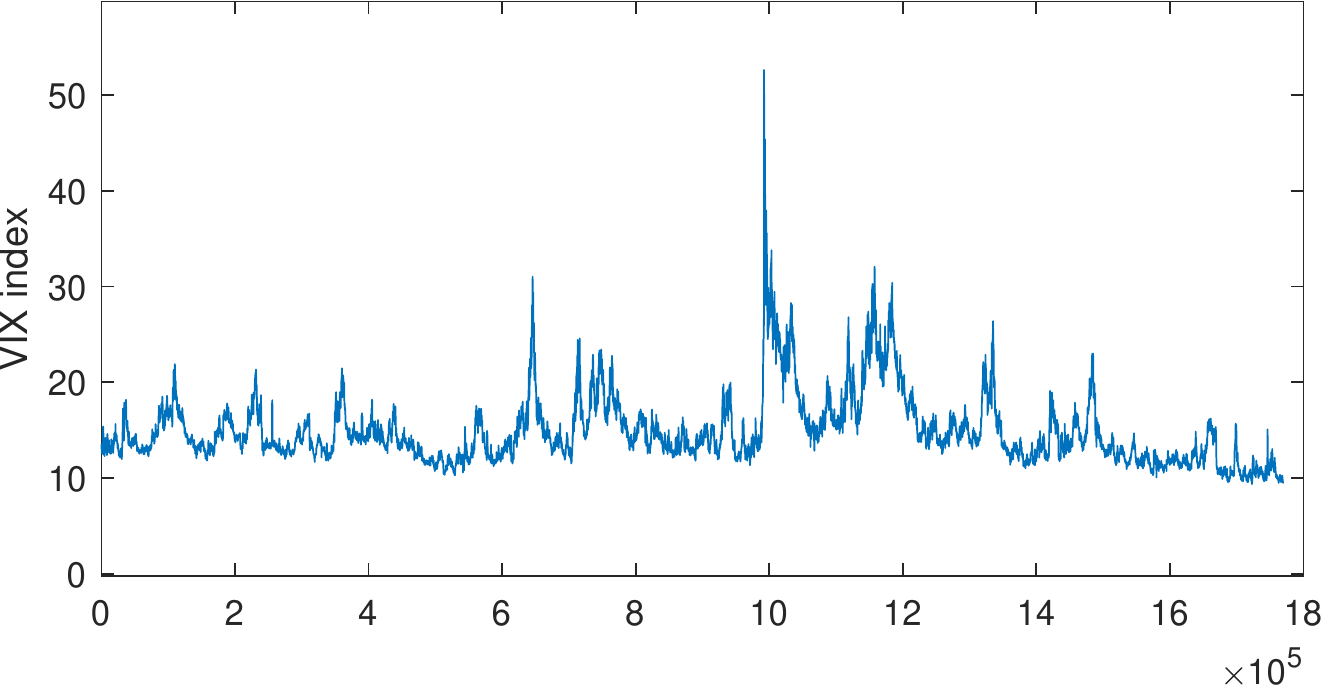}
\caption{\small{Plot of 15-second VIX index from 2013-03-19 to 2017-07-21.}}
\label{fig:fig1}
\end{figure}

The data set we considered records the 15-second VIX index from 2013-03-19 to 2017-07-21, where 15 seconds is the highest frequency available for the VIX index (Chicago Board Option Exchange). The VIX index of the investigation period is plotted in Figure~\ref{fig:fig1}. It is worth noting that the timing of the first and last VIX records can vary slightly on different trading days. To ensure the start time and end time of the daily curves of the VIX records are constant, we use linear interpolation to fill in missing values (if any) so that the timings of the VIX indexes are the same for every trading day. After linear interpolation, we have a total of 1095 trading days (excluding weekends and holidays) in our investigated data. On each trading day, VIX indexes take from 09:31:10 to 16:15:00 of a 15-s interval, and the total constitutes 1616 points per day. 

\begin{figure}[!htb]
\centering
\includegraphics[width=15cm]{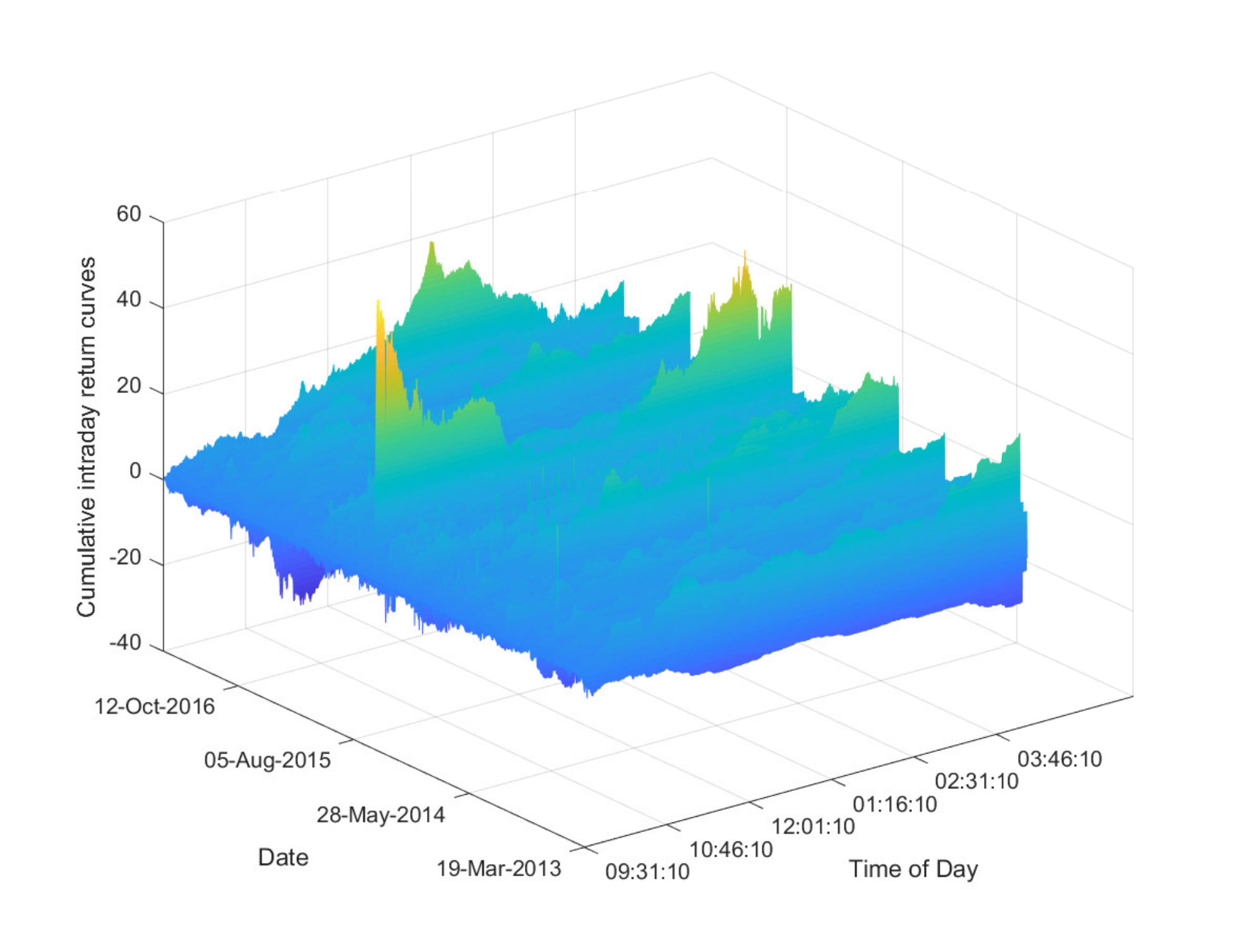}
\caption{\small{Graphical display of functional time series curves of CIDR VIX index from 2013-03-19 to 2017-07-21.}}
\label{fig:fig2}
\end{figure}

Based on the interpolated index, we transformed the non-stationary intraday VIX index into daily curves of cumulative intraday returns (CIDR). Let $P_{i}(t_{j})$ denote the daily VIX value at time $t_{j}$ ($j=1,\ldots,m$) on day $i$ ($i=1,\ldots,n$), CIDRs are computed by 
\begin{equation}
R_{i}(t_{j})=100\times \left[\ln P_{i}(t_{j})-\ln P_{i}(t_{1}) \right],
\end{equation}
where $\ln(\dot)$ represent the natural logarithm and $t_{j-1}$ and $t_{j}$ are 15 seconds apart. The daily curves of the CIDR of the VIX index are the functional time series of interest. Figure~\ref{fig:fig2} provides the plot of the functional time series curves of the CIDR VIX index over different trading days. 

\begin{figure}[!htb]
\centering
\subfloat[A plot of fitted daily CIDR VIX index curves from fAR(1) model.]
{\includegraphics[width=8.3cm]{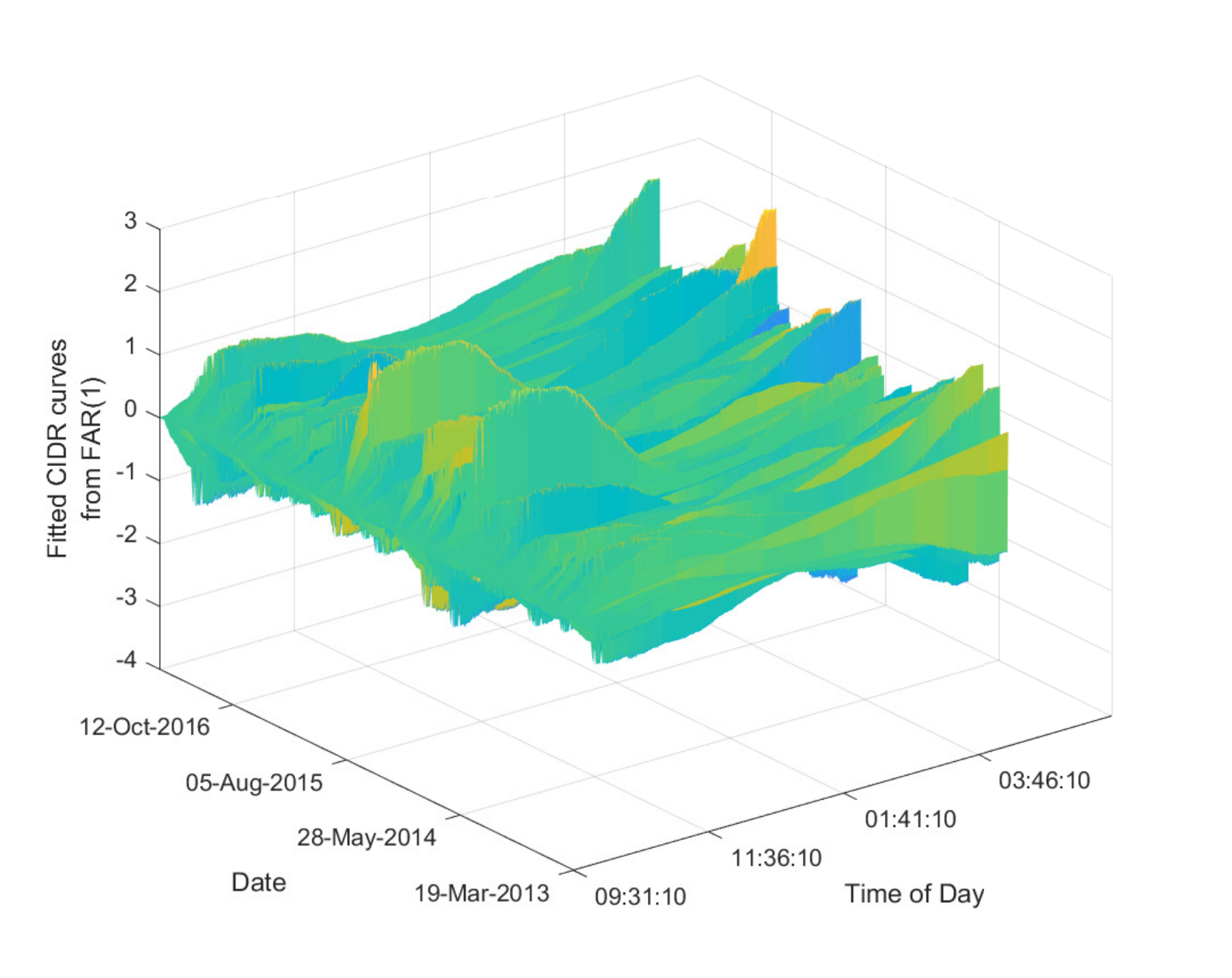} \label{fig:Fig_31}}
\qquad
\subfloat[A plot of fitted conditional  standard deviation from fGARCH(1,1) model.]
{\includegraphics[width=8.3cm]{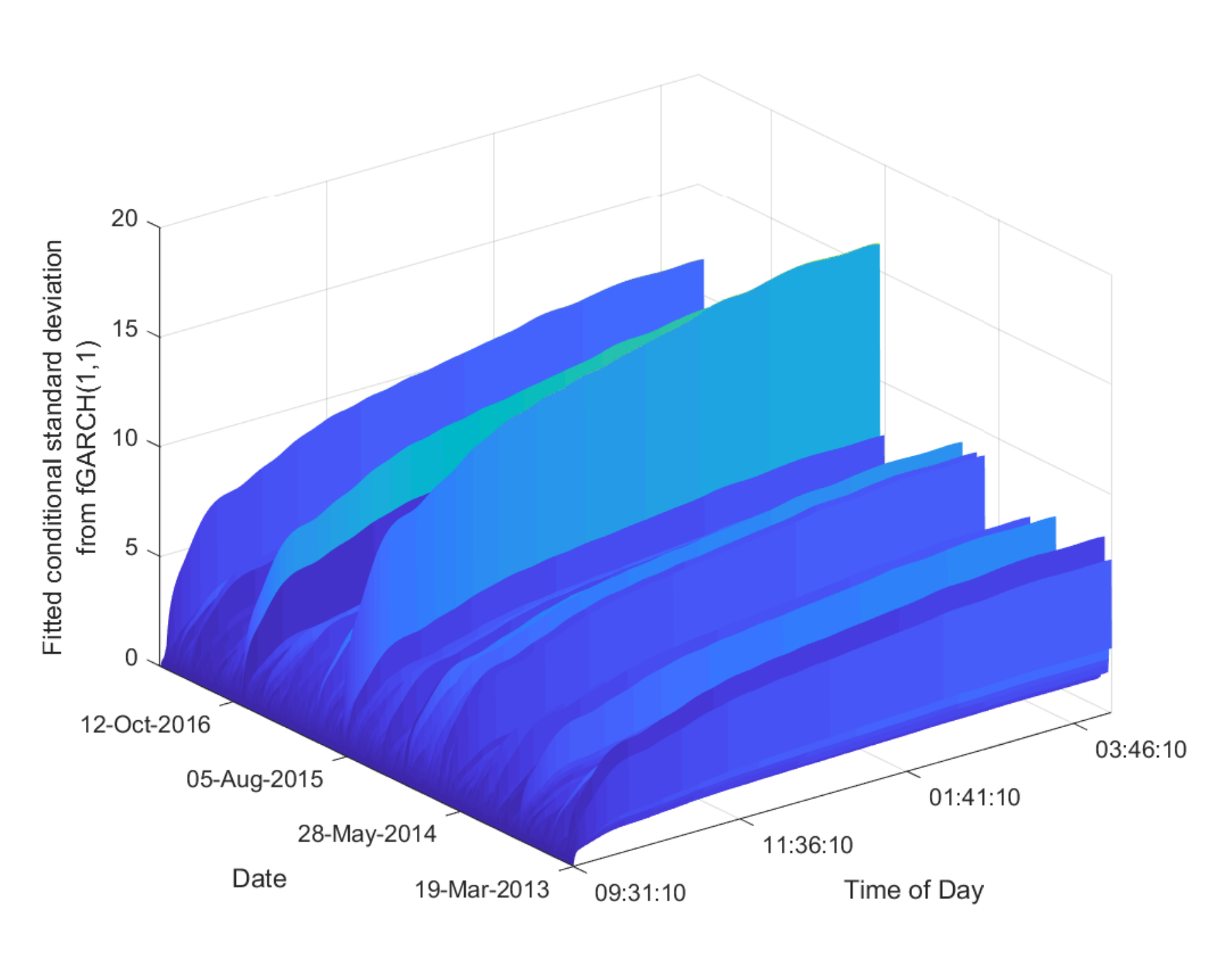}\label{fig:Fig_32}}
\caption{\small{Plots of fitted functional time series curves of CIDR VIX index from fAR(1) model and fitted conditional  standard deviation ($\widehat{\sigma}_{i}$) from fGARCH$(1,1)$ model.}}\label{fig:fig3}
\end{figure}

The candidate models we consider to fit the daily curves of the CIDR VIX index are the fAR(1) model and fGARCH(1,1) model. The specification of the fAR(1) model is provided in~\eqref{FAR1} of Section~\ref{section4}. We use the package 'far' to fit the observed data to the fAR(1) model. The fGARCH model is proposed by \cite{aue2017functional}. A sequence of random functions $(R_{i}:i\in \mathbb{Z})$ is called a fGARCH(1,1), if it satisfies the equations
\begin{align}\label{fgarch1}
R_{i}&=\sigma_{i}\varepsilon_{i},\\
\sigma^{2}_{i}&=\delta+\alpha R^{2}_{i-1}+\beta \sigma^{2}_{i-1},
\end{align}  
where $\delta$ is a non-negative function, the operators $\alpha$ and $\beta$ map non-negative functions to non-negative functions and the innovations $\varepsilon_{i}$ are IID random functions. The estimation procedure for fitting the fGARCH model is described in \cite{RWZ21} via quasi-likelihood. The fitted VIX index returns estimated from the fAR(1) model and the fitted $\sigma_{i}$ estimated from the fGARCH(1,1) model are plotted in Figure~\ref{fig:fig3}.

To evaluate the adequacy of the fAR(1) model, we apply the functional BDS test on the residuals between the observed returns curves and the fitted return curves ($(R_{i}(t_{j})-\widehat{R}_{i}(t_{j}):j=1,\ldots,\tau;i=1,\ldots,n)$). Table~\ref{tb:BDSE1} presents the functional BDS test statistics of fAR(1) model residuals for a variety combination of hyper-parameter $m$ and $r$. Since most test statistics exceed the 1\% critical value of standard normal distribution, the functional BDS test rejects the null hypothesis of IID residuals. In other words, the fAR(1) model cannot capture all structures underlying the observed daily curves of VIX returns.

\begin{table}[!htbp]
\tabcolsep 0.195in
\centering
\caption{\small{The functional BDS test statistics of fAR(1) residuals and fGARCH(1,1) logarithm squared standardized returns fitted to daily curves of the CIDR VIX index. * indicates the independent null hypothesis is rejected at 5\% significance, and ** indicates rejection at 1\% significance.}}\label{tab:BDS_VIX}
\begin{tabular}{@{}llrrrrrr@{}}
\toprule  
			&				& \multicolumn{6}{c}{$m$} \\
Model		& $r$ 			&  $2$ &  $3$ &  $4$ &  $5$ &  $6$ & $7$ \\ \midrule
fAR(1) 		& $\text{s.d.}$ 		& 7.75** & 9.15** & 9.52** & 10.05** & 10.39** 	& 11.11**\\
			& $1.25\text{s.d.}$ 	& 6.26** & 7.73** & 8.20** & 8.57**  	& 8.71**  	& 9.15** \\
			& $1.5\text{s.d.}$ 	& 5.05** & 6.67** & 7.21** & 7.68**  	& 7.76**  	& 8.13** \\
 \\
fGARCH(1,1)  		& $\text{s.d.}$ 		& 1.42 	& 2.06* 	& 2.77** & 3.25** & 3.66** & 3.93**\\
& $1.25\text{s.d.}$ 	& 2.32*& 2.69** 	& 3.09** 	& 3.21** 	& 3.47** & 3.73**   \\
				& $1.5\text{s.d.}$ 	& 2.72** 	& 3.13** 	& 3.42** & 3.42** & 3.59** & 3.75**\\ 
\bottomrule
\end{tabular}
\label{tb:BDSE1}
\end{table}
 
Since the fGARCH(1,1) is a multiplicative model, we use the standardized returns ($R_{i}/\widehat{\sigma}_{i}$) to evaluate its adequacy. In the univariate case,  when evaluating the adequacy of the GARCH model, if the BDS test is applied directly to the standardized returns $R_{t}/\widehat{\sigma}_{t}$, earlier studies \citep[see][]{brock1991nonlinear} suggest the BDS statistic needs to be adjusted to have the right size. \cite{fernandes2012finite} proposed to apply the BDS test on natural logarithms of squared standardized residuals ($\ln(R_{t}^{2}/\widehat{\sigma}_{t}^{2})$) so that the logarithmic transformation casts the GARCH model into a linear additive model. Table~\ref{tb:BDSE1} records the functional BDS test statistics of the logarithm of the squared standardized returns. 

To compare the newly proposed BDS test with the existing independence test in the functional time series literature, we conduct the independence test proposed by \cite{gabrys2007portmanteau} (abbreviated as the GK independence test) on the fAR(1) residuals and fGARCH(1,1) standardized returns. The GK independence test statistic is based on the lagged cross-covariances of the projected principal components of the functional time series. Therefore, the GK independence test statistic is restricted in detecting linear structure. The GK independence test requires two hyperparameters, $p$ and $H$. The hyperparameter $p$ represents the number of retained principal components in the dimension reduction step, and $H$ denotes the maximum lagged cross-covariances considered in computing the test statistics. Table~\ref{tb:GKE} documents the $p$-value of the GK independence test for $H=1, 10$ and $p=3, 4, 5, 8, 10$. In \cite{gabrys2007portmanteau}, the authors investigated the finite-sample performance of the GK independence test with $H=1,3,5$ and $p=3,4,5$ and concluded that the test power against the fAR(1) model is very good if $H=1$ is used. Since the optimal parameters of the GK test depend on the underlying dynamic of the residuals, which is unknown in empirical studies, we also extend the GK test with relatively larger $H$ and $p$.

\begin{table}[!htbp]
\tabcolsep 0.3in
\centering
\caption{\small{The GK test $p$-value of fAR$(1)$ residuals and fGARCH$(1,1)$ standardized returns fitted to daily curves of the CIDR VIX index. * indicates the independent null hypothesis is rejected at 5\% significance, and ** indicates rejection at 1\% significance.}}\label{tab:GK_VIX}
\begin{tabular}{@{}llrrrrr@{}}
\toprule  
			&				& \multicolumn{5}{c}{$p$} \\
Model		& $H$ 		& $3$ &  $4$	&  $5$ 		&  $8$ 		&  $10$  \\ \midrule
fAR(1)		& 1 	 & 1 & 1  & 1 & 0.36			& 1.40e-05**  \\
 		    & 10 	& 0.70 & 0.58		&  0.29   	& 1.01e-4**  	&  6.43e-11**\\
 		    \\
 		    
 fGARCH(1,1)  	& 1 & 0.37 & 		0.41		&  	0.60	&  0.47		& 0.54 \\

  	& 10  & 0.92 &		0.81		& 0.42 		& 0.78 		& 0.55 \\
\bottomrule
\end{tabular}
\label{tb:GKE}
\end{table}

Comparing the inferences drawn from the functional BDS test and the GK test provides additional insights into the dynamics of the CIDR VIX functional time series. Based on the functional BDS test results, both the fAR(1) and fGARCH(1,1) models are insufficient to capture the temporal structure exhibited in the daily CIDR curves of the VIX index. However, the GK test showed evidence of a violation of independence only for the fAR(1) model when a larger $p$ is selected. This indicates that the fGARCH(1,1) model better fits the observed curves compared to the fAR(1) model. Additionally, the seemingly contradictory conclusions regarding the fGARCH(1,1) residuals from the functional BDS test and the GK test indicate a nonlinear structure exhibited in the daily curves of the CIDR VIX index. Furthermore, the GK test results indicate that the test's inference can vary based on the parameters selected, whereas our functional BDS test provides consistent inferences across different parameter selections.

\section{Conclusion}\label{section6}

In this paper, we extended the BDS test to functional time series. Just like the BDS test in the univariate case, the functional BDS test enjoys some key desired properties, making it a plausible candidate for testing model specification and nonlinearity. Those advantages include a minimal requirement of prior assumptions and knowledge and the capacity to detect linear and nonlinear structures. We proved that the asymptotic normality previously held for the test statistics under the null hypothesis in the univariate case stays valid after adjusting the test statistics to the functional case. Additionally, we conducted Monte-Carlo experiments on the functional BDS test to provide the recommended range of its hyperparameters and data length. We showed that with the appropriate selection of the hyperparameters, the functional BDS test only required the data to be of length 250 to ensure its converges to normality and have a 100\% correct rate in detecting the predictability in a simulated functional time series with a relatively weak temporal structure. Moreover, if either $L_{1}$  or $L_{2}$ is selected as the distance measure inside the sup-norms, the function BDS test is also robust to outliers. The code for the functional BDS test is available at \url{https://github.com/Landy339/functional_BDS_test}.

We illustrate the significance of our research in an empirical analysis where we used the functional BDS test to evaluate the adequacy of the fAR(1) model and the fGARCH(1,1) model in fitting the CIDR VIX index functional time series. After fitting the candidate models, we applied the functional BDS test to detect the remaining structures in the residuals. The test rejects the independence null hypothesis and thus concludes that both fAR(1) and fGARCH(1,1) models are insufficient to fully capture the temporal structures exhibited in the observed curves. Additionally, our test showed added sensitivity in detecting predictability, especially the nonlinear structure, compared to the existing independence test in functional time series. We compared the results from the functional BDS test with that from the GK test, an existing linear independent test in the domain of functional time series. The results showed that our newly proposed functional BDS test remedies the weakness of the GK test by detecting the nonlinear structure in the fGARCH(1,1) residuals that the GK test neglects. With the new tool, one could be aware of the existing independence test that the fGARCH(1,1) is an adequate model for the observed data and overlook its nonlinear temporal structures.

The functional BDS test is the first nonlinearity test and the first model specification test proposed in functional time series. However, the major limitation of the proposed test is that it can only detect the remaining structures in the residuals. Unfortunately, it cannot indicate the form of the detected structures. Consequently, if a model is deemed insufficient, practitioners have no guidance on what kind of models can fully capture the structures in the observed data.

There are several ways in which the current work may be further extended, and we briefly outline three possibilities: 
\begin{inparaenum}
\item[(1)] Although our study demonstrated that with the proper selection of norms, the functional BDS test is robust to outliers, future research can examine its behavior on non-stationary functional time series, which frequently arise in real-world data. 
\item[(2)] The current study focused on univariate functional time series. Future work could investigate the extension of nonlinearity tests to multivariate functional time series while accounting for potential correlations among the variables. 
\item[(3)] Since our empirical analysis indicates the existence of nonlinearity in financial functional time series, we hope that our work will inspire further research into the dependence structure of functional time series, particularly in analyzing, modeling, and forecasting nonlinear functional time series.
\end{inparaenum}

\section*{Acknowledgments}

This work was supported by the Australian Research Council under Grant DP230102250 titled ``Feature learning for high-dimensional functional time series". The authors would like to thank Dr. Yuqian Zhao for generously sharing his code used for part of the empirical studies. 

\newpage
\appendix
\section{Appendix: Additional Monte-Carlo simulation results}\label{appendix1}

In Table~\ref{tb:BDS1}, we present the $p$-value of the KS test on the functional BDS test statistics computed on simulated IID functional time series with $L_{1}$ norm and $L_{\text{inf}}$ norm being used as the distance measure inside the sup-norms.

\begin{table}[!htbp]
\tabcolsep 0.17in
\centering
\caption{\small{The $p$-value of the KS test on functional BDS test statistics with $L_{1}$ norm and $L_{\text{inf}}$ norm computed on 200 paths of 500 simulated IID functional time series.}}
\begin{tabular}{@{}llrrrrrrrrr@{}}
\toprule  
		& & \multicolumn{9}{c}{$m$} \\
Metric &	 $r$ 	&  $2$ &  $3$ &  $4$ &  $5$ &  $6$ &  $7$ & $8$ &  $9$ &  $10$\\ \midrule
$L_1$  &
  $0.25\text{s.d.}$ &
  {\color[HTML]{FE0000} 0.00} &
  {\color[HTML]{FE0000} 0.00} &
  {\color[HTML]{FE0000} 0.00} &
  {\color[HTML]{FE0000} 0.00} &
  {\color[HTML]{FE0000} 0.00} &
  {\color[HTML]{FE0000} 0.00} &
  {\color[HTML]{FE0000} 0.00} &
  {\color[HTML]{FE0000} 0.00} &
  {\color[HTML]{FE0000} 0.00} \\
 &
  $0.5\text{s.d.}$ &
  {\color[HTML]{FE0000} 0.00} &
  {\color[HTML]{FE0000} 0.00} &
  {\color[HTML]{FE0000} 0.00} &
  {\color[HTML]{FE0000} 0.00} &
  {\color[HTML]{FE0000} 0.00} &
  {\color[HTML]{FE0000} 0.00} &
  {\color[HTML]{FE0000} 0.00} &
  {\color[HTML]{FE0000} 0.00} &
  {\color[HTML]{FE0000} 0.00} \\
 &
  $0.75\text{s.d.}$ &
  {\color[HTML]{FE0000} 0.00} &
  {\color[HTML]{FE0000} 0.46} &
  {\color[HTML]{FE0000} 0.04} &
  {\color[HTML]{FE0000} 0.00} &
  {\color[HTML]{FE0000} 0.00} &
  {\color[HTML]{FE0000} 0.00} &
  {\color[HTML]{FE0000} 0.00} &
  {\color[HTML]{FE0000} 0.00} &
  {\color[HTML]{FE0000} 0.00} \\
 &
  $\text{s.d.}$ &
  0.78 &
  0.55 &
  0.14 &
  0.04 &
  0.08 &
  0.17 &
  0.03 &
  0.03 &
  {\color[HTML]{FE0000} 0.00} \\
 &
 $1.25\text{s.d.}$ &
  0.89 &
  0.19 &
  0.94 &
  0.26 &
  {\color[HTML]{FE0000} 0.00} &
  0.03 &
  {\color[HTML]{FE0000} 0.00} &
  {\color[HTML]{FE0000} 0.00} &
  {\color[HTML]{FE0000} 0.00} \\
 &
 $1.5\text{s.d.}$ &
  0.78 &
  0.15 &
  0.03 &
  0.03 &
  0.22 &
  {\color[HTML]{FE0000} 0.01} &
  0.03 &
  {\color[HTML]{FE0000} 0.00} &
  {\color[HTML]{FE0000} 0.00} \\
 &
 $1.75\text{s.d.}$ &
  0.24 &
  0.30 &
  0.74 &
  0.15 &
  {\color[HTML]{FE0000} 0.00} &
  0.57 &
  0.08 &
  {\color[HTML]{FE0000} 0.02} &
  0.27 \\
 &
  $2\text{s.d.}$ &
  0.54 &
  0.09 &
  0.21 &
  0.09 &
  {\color[HTML]{FE0000} 0.01} &
  0.10 &
  {\color[HTML]{FE0000} 0.01} &
  0.22 &
  0.03      \\ 
\\
$L_{\text{inf}}$ &
  $0.25\text{s.d.}$ &
  {\color[HTML]{FE0000} 0.00} &
  {\color[HTML]{FE0000} 0.00} &
  {\color[HTML]{FE0000} 0.00} &
  {\color[HTML]{FE0000} 0.00} &
  {\color[HTML]{FE0000} 0.00} &
  {\color[HTML]{FE0000} 0.00} &
  {\color[HTML]{FE0000} 0.00} &
  {\color[HTML]{FE0000} 0.00} &
  {\color[HTML]{FE0000} 0.00} \\
 &
  $0.5\text{s.d.}$ &
  0.09 &
  {\color[HTML]{FE0000} 0.00} &
  {\color[HTML]{FE0000} 0.00} &
  {\color[HTML]{FE0000} 0.00} &
  {\color[HTML]{FE0000} 0.00} &
  {\color[HTML]{FE0000} 0.00} &
  {\color[HTML]{FE0000} 0.00} &
  {\color[HTML]{FE0000} 0.00} &
  {\color[HTML]{FE0000} 0.00} \\
 &
  $0.75\text{s.d.}$ &
  0.10 &
  0.19 &
  0.29 &
  {\color[HTML]{FE0000} 0.01} &
  {\color[HTML]{FE0000} 0.00} &
  {\color[HTML]{FE0000} 0.00} &
  {\color[HTML]{FE0000} 0.00} &
  {\color[HTML]{FE0000} 0.00} &
  {\color[HTML]{FE0000} 0.00} \\
 &
  $\text{s.d.}$ &
  {\color[HTML]{FE0000} 0.00} &
  0.53 &
  0.82 &
  0.20 &
  0.03 &
  0.03 &
  {\color[HTML]{FE0000} 0.00} &
  {\color[HTML]{FE0000} 0.00} &
  {\color[HTML]{FE0000} 0.00} \\
 &
  $1.25\text{s.d.}$ &
  0.41 &
  0.83 &
  0.98 &
  0.58 &
  {\color[HTML]{FE0000} 0.01} &
  0.11 &
  {\color[HTML]{FE0000} 0.00} &
  0.34 &
  {\color[HTML]{FE0000} 0.00} \\
 &
  $1.5\text{s.d.}$ &
  0.72 &
  0.29 &
  0.72 &
  0.17 &
  0.12 &
  0.25 &
  0.09 &
  {\color[HTML]{FE0000} 0.02} &
  {\color[HTML]{FE0000} 0.01} \\
 &
 $1.75\text{s.d.}$ &
  0.86 &
  0.03 &
  0.53 &
  0.40 &
  0.22 &
  {\color[HTML]{FE0000} 0.02} &
  0.35 &
  0.11 &
  {\color[HTML]{FE0000} 0.00} \\
 &
  $2\text{s.d.}$ &
  {\color[HTML]{FE0000} 0.00} &
  0.16 &
  0.29 &
  0.13 &
  0.91 &
  0.17 &
  0.22 &
  0.93 &
  {\color[HTML]{FE0000} 0.00}   \\
\bottomrule
\end{tabular}
\label{tb:BDS1}
\end{table}

In Table~\ref{tb:BDSN1}, we present the probability that the functional BDS test successfully rejects the IID hypothesis on a structured process when $L_1$ and $L_{\text{inf}}$ are selected as the norms.

\begin{table}[!htbp]
\tabcolsep 0.125in
\centering
\caption{\small{The successful rejection rate of the functional BDS test on 200 paths of the simulated fAR(1) process ($\rho=0.1$) of 500 observations with $L_{1}$ and $L_{\text{inf}}$ being used as the norm inside the sup-norms and different choices of $m$ and $r$.}}
\begin{tabular}{@{}llrrrrrrrrr@{}}
\toprule  
		& & \multicolumn{9}{c}{$m$} \\
Metric &	 $r$ 	&  $2$ &  $3$ &  $4$ &  $5$ &  $6$ &  $7$ & $8$ &  $9$ &  $10$\\ \midrule
$L_1$ & $0.25\text{s.d.}$ &
  100\% &
  {\color[HTML]{FE0000} 98\%} &
  100\% &
  100\% &
  {\color[HTML]{FE0000} 98\%} &
  {\color[HTML]{FE0000} 73\%} &
  {\color[HTML]{FE0000} 19\%} &
  {\color[HTML]{FE0000} 1\%} &
  {\color[HTML]{FE0000} 1\%}   \\
& $0.5\text{s.d.}$     & 100\% & 100\% & 100\% & 100\% & 100\% & 100\% & {\color[HTML]{FE0000} 93\%} & {\color[HTML]{FE0000} 51\%} & {\color[HTML]{FE0000} 15\%}     \\
	& $0.75\text{s.d.}$  & 100\% & 100\% & 100\% & 100\% & 100\% & 100\% & 100\%                       & {\color[HTML]{FE0000} 99\%} & {\color[HTML]{FE0000} 97\%}    \\
	& $\text{s.d.}$   & 100\% & 100\% & 100\% & 100\% & 100\% & 100\% & 100\%                       & 100\%                       & 100\%        \\
	& $1.25\text{s.d.}$   & 100\% & 100\% & 100\% & 100\% & 100\% & 100\% & 100\%                       & 100\%                       & 100\%       \\
	& $1.5\text{s.d.}$   &
  100\% &
  {\color[HTML]{FE0000} 100\%} &
  100\% &
  {\color[HTML]{FE0000} 100\%} &
  {\color[HTML]{FE0000} 98\%} &
  100\% &
  100\% &
  {\color[HTML]{FE0000} 99\%} &
  {\color[HTML]{FE0000} 98\%}    \\
	& $1.75\text{s.d.}$  &
  {\color[HTML]{FE0000} 96\%} &
  {\color[HTML]{FE0000} 98\%} &
  {\color[HTML]{FE0000} 98\%} &
  {\color[HTML]{FE0000} 96\%} &
  {\color[HTML]{FE0000} 95\%} &
  {\color[HTML]{FE0000} 96\%} &
  {\color[HTML]{FE0000} 94\%} &
  {\color[HTML]{FE0000} 91\%} &
  {\color[HTML]{FE0000} 91\%}     \\
	& $2\text{s.d.}$ &
  {\color[HTML]{FE0000} 91\%} &
  {\color[HTML]{FE0000} 93\%} &
  {\color[HTML]{FE0000} 85\%} &
  {\color[HTML]{FE0000} 84\%} &
  {\color[HTML]{FE0000} 89\%} &
  {\color[HTML]{FE0000} 87\%} &
  {\color[HTML]{FE0000} 86\%} &
  {\color[HTML]{FE0000} 84\%} &
  {\color[HTML]{FE0000} 78\%}      \\ 
\\
$L_{\text{inf}}$ &
  $0.25\text{s.d.}$ &
  100\% &
  100\% &
  100\% &
  100\% &
  {\color[HTML]{FE0000} 94\%} &
  {\color[HTML]{FE0000} 51\%} &
  {\color[HTML]{FE0000} 11\%} &
  {\color[HTML]{FE0000} 1\%} &
  {\color[HTML]{FE0000} 1\%} \\
 &
  $0.5\text{s.d.}$ &
  100\% &
  100\% &
  100\% &
  100\% &
  100\% &
  100\% &
  {\color[HTML]{FE0000} 83\%} &
  {\color[HTML]{FE0000} 35\%} &
  {\color[HTML]{FE0000} 8\%} \\
 & $0.75\text{s.d.}$                  & 100\% & 100\% & 100\% & 100\% & 100\% & 100\% & 100\% & 100\% & {\color[HTML]{FE0000} 94\%} \\
 & $\text{s.d.}$                       & 100\% & 100\% & 100\% & 100\% & 100\% & 100\% & 100\% & 100\% & 100\%                       \\
 & $1.25\text{s.d.}$                   & 100\% & 100\% & 100\% & 100\% & 100\% & 100\% & 100\% & 100\% & 100\%                       \\
 & $1.5\text{s.d.}$                  & 100\% & 100\% & 100\% & 100\% & 100\% & 100\% & 100\% & 100\% & 100\%                       \\
 & $1.75\text{s.d.}$                  & 100\% & 100\% & 100\% & 100\% & 100\% & 100\% & 100\% & 100\% & 100\%                       \\
 &
  $2\text{s.d.}$ &
  {\color[HTML]{FE0000} 99\%} &
  100\% &
  100\% &
  {\color[HTML]{FE0000} 99\%} &
  {\color[HTML]{FE0000} 99\%} &
  {\color[HTML]{FE0000} 99\%} &
  {\color[HTML]{FE0000} 99\%} &
  {\color[HTML]{FE0000} 98\%} &
  {\color[HTML]{FE0000} 99\%}  \\
\bottomrule
\end{tabular}
\label{tb:BDSN1}
\end{table}

Table~\ref{tb:sensitivity} stores the p-value of the KS test on the functional BDS test statistics computed on simulated IID functional time series with 1\% random outliers when $L_{1}$, $L_{2}$ and $L_{\text{inf}}$ are used as the norm inside the sup-norms.

\begin{table}[!htbp]
\tabcolsep 0.17in
\centering
\caption{\small{The $p$-value of the KS test on functional BDS test statistics with $L_{1}$, $L_{2}$ and $L_{\text{inf}}$ norm computed on 200 paths of 500 simulated IID functional time series with 1\% random outliers.}}
\begin{tabular}{@{}llrrrrrrrrr@{}}
\toprule  
		& & \multicolumn{9}{c}{$m$} \\
Metric &	 $r$ 	&  $2$ &  $3$ &  $4$ &  $5$ &  $6$ &  $7$ & $8$ &  $9$ &  $10$\\ \midrule
$L_1$ &
  $0.25\text{s.d.}$ &
  {\color[HTML]{FE0000} 0.00} &
  {\color[HTML]{FE0000} 0.00} &
  {\color[HTML]{FE0000} 0.00} &
  {\color[HTML]{FE0000} 0.00} &
  {\color[HTML]{FE0000} 0.00} &
  {\color[HTML]{FE0000} 0.00} &
  {\color[HTML]{FE0000} 0.00} &
  {\color[HTML]{FE0000} 0.00} &
  {\color[HTML]{FE0000} 0.00} \\
 &
  $0.5\text{s.d.}$ &
  0.05 &
  {\color[HTML]{FE0000} 0.00} &
  {\color[HTML]{FE0000} 0.00} &
  {\color[HTML]{FE0000} 0.00} &
  {\color[HTML]{FE0000} 0.00} &
  {\color[HTML]{FE0000} 0.00} &
  {\color[HTML]{FE0000} 0.00} &
  {\color[HTML]{FE0000} 0.00} &
  {\color[HTML]{FE0000} 0.00} \\
 &
  $0.75\text{s.d.}$ &
  0.33 &
  0.05 &
  {\color[HTML]{FE0000} 0.00} &
  {\color[HTML]{FE0000} 0.00} &
  {\color[HTML]{FE0000} 0.00} &
  {\color[HTML]{FE0000} 0.00} &
  {\color[HTML]{FE0000} 0.00} &
  {\color[HTML]{FE0000} 0.00} &
  {\color[HTML]{FE0000} 0.00} \\
 &
  $\text{s.d.}$ &
  0.68 &
  0.25 &
  0.39 &
  0.25 &
  0.03 &
  {\color[HTML]{FE0000} 0.00} &
  {\color[HTML]{FE0000} 0.00} &
  {\color[HTML]{FE0000} 0.00} &
  {\color[HTML]{FE0000} 0.00} \\
 &
  $1.25\text{s.d.}$ &
  0.36 &
  0.37 &
  {\color[HTML]{FE0000} 0.02} &
  0.16 &
  0.03 &
  0.17 &
  {\color[HTML]{FE0000} 0.02} &
  0.12 &
  {\color[HTML]{FE0000} 0.00} \\
 &
  $1.5\text{s.d.}$ &
  0.80 &
  0.12 &
  0.93 &
  0.09 &
  0.15 &
  0.43 &
  0.98 &
  0.02 &
  0.09 \\
 &
  $1.75\text{s.d.}$ &
  0.40 &
  0.70 &
  0.76 &
  0.16 &
  0.85 &
  0.21 &
  0.49 &
  0.42 &
  {\color[HTML]{FE0000} 0.01} \\
 &
  $2\text{s.d.}$ &
  0.83 &
  0.17 &
  0.13 &
  {\color[HTML]{FE0000} 0.02} &
  0.28 &
  0.05 &
  0.26 &
  0.67 &
  0.08 \\
 &
   &
   &
   &
   &
   &
   &
   &
   &
   &
   \\
$L_2$ &
  $0.25\text{s.d.}$ &
  {\color[HTML]{FE0000} 0.00} &
  {\color[HTML]{FE0000} 0.00} &
  {\color[HTML]{FE0000} 0.00} &
  {\color[HTML]{FE0000} 0.00} &
  {\color[HTML]{FE0000} 0.00} &
  {\color[HTML]{FE0000} 0.00} &
  {\color[HTML]{FE0000} 0.00} &
  {\color[HTML]{FE0000} 0.00} &
  {\color[HTML]{FE0000} 0.00} \\
 &
  $0.5\text{s.d.}$ &
  {\color[HTML]{FE0000} 0.01} &
  {\color[HTML]{FE0000} 0.00} &
  {\color[HTML]{FE0000} 0.00} &
  {\color[HTML]{FE0000} 0.00} &
  {\color[HTML]{FE0000} 0.00} &
  {\color[HTML]{FE0000} 0.00} &
  {\color[HTML]{FE0000} 0.00} &
  {\color[HTML]{FE0000} 0.00} &
  {\color[HTML]{FE0000} 0.00} \\
 &
  $0.75\text{s.d.}$ &
  0.33 &
  {\color[HTML]{FE0000} 0.00} &
  {\color[HTML]{FE0000} 0.00} &
  {\color[HTML]{FE0000} 0.00} &
  {\color[HTML]{FE0000} 0.00} &
  {\color[HTML]{FE0000} 0.00} &
  {\color[HTML]{FE0000} 0.00} &
  {\color[HTML]{FE0000} 0.00} &
  {\color[HTML]{FE0000} 0.00} \\
 &
  $\text{s.d.}$ &
  0.67 &
  0.30 &
  0.08 &
  {\color[HTML]{FE0000} 0.00} &
  {\color[HTML]{FE0000} 0.00} &
  {\color[HTML]{FE0000} 0.00} &
  {\color[HTML]{FE0000} 0.00} &
  {\color[HTML]{FE0000} 0.00} &
  {\color[HTML]{FE0000} 0.00} \\
 &
  $1.25\text{s.d.}$ &
  0.62 &
  0.23 &
  0.10 &
  0.23 &
  0.15 &
  {\color[HTML]{FE0000} 0.00} &
  {\color[HTML]{FE0000} 0.00} &
  {\color[HTML]{FE0000} 0.00} &
  {\color[HTML]{FE0000} 0.00} \\
 &
  $1.5\text{s.d.}$ &
  0.57 &
  0.95 &
  0.07 &
  0.15 &
  0.35 &
  0.05 &
  {\color[HTML]{FE0000} 0.00} &
  0.11 &
  {\color[HTML]{FE0000} 0.00} \\
 &
  $1.75\text{s.d.}$ &
  0.18 &
  0.35 &
  0.49 &
  0.19 &
  {\color[HTML]{FE0000} 0.01} &
  0.31 &
  0.03 &
  {\color[HTML]{FE0000} 0.01} &
  0.23 \\
 &
  $2\text{s.d.}$ &
  {\color[HTML]{FE0000} 0.01} &
  0.16 &
  0.16 &
  0.03 &
  {\color[HTML]{FE0000} 0.02} &
  0.19 &
  0.33 &
  0.26 &
  {\color[HTML]{FE0000} 0.01} \\
 &
   &
   &
   &
   &
   &
   &
   &
   &
   &
   \\
$L_{\text{inf}}$ &
  $0.25\text{s.d.}$ &
  {\color[HTML]{FE0000} 0.00} &
  {\color[HTML]{FE0000} 0.00} &
  {\color[HTML]{FE0000} 0.00} &
  {\color[HTML]{FE0000} 0.00} &
  {\color[HTML]{FE0000} 0.00} &
  {\color[HTML]{FE0000} 0.00} &
  {\color[HTML]{FE0000} 0.00} &
  {\color[HTML]{FE0000} 0.00} &
  {\color[HTML]{FE0000} 0.00} \\
 &
  $0.5\text{s.d.}$ &
  {\color[HTML]{FE0000} 0.00} &
  {\color[HTML]{FE0000} 0.00} &
  {\color[HTML]{FE0000} 0.00} &
  {\color[HTML]{FE0000} 0.00} &
  {\color[HTML]{FE0000} 0.00} &
  {\color[HTML]{FE0000} 0.00} &
  {\color[HTML]{FE0000} 0.00} &
  {\color[HTML]{FE0000} 0.00} &
  {\color[HTML]{FE0000} 0.00} \\
 &
  $0.75\text{s.d.}$ &
  0.44 &
  {\color[HTML]{FE0000} 0.00} &
  {\color[HTML]{FE0000} 0.00} &
  {\color[HTML]{FE0000} 0.00} &
  {\color[HTML]{FE0000} 0.00} &
  {\color[HTML]{FE0000} 0.00} &
  {\color[HTML]{FE0000} 0.00} &
  {\color[HTML]{FE0000} 0.00} &
  {\color[HTML]{FE0000} 0.00} \\
 &
  $\text{s.d.}$ &
  0.71 &
  0.09 &
  {\color[HTML]{FE0000} 0.00} &
  {\color[HTML]{FE0000} 0.00} &
  {\color[HTML]{FE0000} 0.00} &
  {\color[HTML]{FE0000} 0.00} &
  {\color[HTML]{FE0000} 0.00} &
  {\color[HTML]{FE0000} 0.00} &
  {\color[HTML]{FE0000} 0.00} \\
 &
  $1.25\text{s.d.}$ &
  0.47 &
  0.54 &
  {\color[HTML]{FE0000} 0.00} &
  {\color[HTML]{FE0000} 0.00} &
  {\color[HTML]{FE0000} 0.00} &
  {\color[HTML]{FE0000} 0.00} &
  {\color[HTML]{FE0000} 0.00} &
  {\color[HTML]{FE0000} 0.00} &
  {\color[HTML]{FE0000} 0.00} \\
 &
  $1.5\text{s.d.}$ &
  0.03 &
  0.71 &
  {\color[HTML]{FE0000} 0.00} &
  {\color[HTML]{FE0000} 0.00} &
  {\color[HTML]{FE0000} 0.00} &
  {\color[HTML]{FE0000} 0.00} &
  {\color[HTML]{FE0000} 0.00} &
  {\color[HTML]{FE0000} 0.00} &
  {\color[HTML]{FE0000} 0.00} \\
 &
  $1.75\text{s.d.}$ &
  0.17 &
  0.08 &
  {\color[HTML]{FE0000} 0.00} &
  {\color[HTML]{FE0000} 0.00} &
  {\color[HTML]{FE0000} 0.00} &
  {\color[HTML]{FE0000} 0.00} &
  {\color[HTML]{FE0000} 0.00} &
  {\color[HTML]{FE0000} 0.00} &
  {\color[HTML]{FE0000} 0.00} \\
 &
  $2\text{s.d.}$ &
  0.35 &
  {\color[HTML]{FE0000} 0.00} &
  {\color[HTML]{FE0000} 0.00} &
  {\color[HTML]{FE0000} 0.00} &
  {\color[HTML]{FE0000} 0.00} &
  {\color[HTML]{FE0000} 0.00} &
  {\color[HTML]{FE0000} 0.00} &
  {\color[HTML]{FE0000} 0.00} &
  {\color[HTML]{FE0000} 0.00}

\\
\bottomrule
\end{tabular}
\label{tb:sensitivity}
\end{table}

\section{Appendix: Asymptotic Normality for the BDS Test Statistic and its Proof}\label{appendix2}

In Section~\ref{section2}, we adjust the specification of the BDS test statistic to be adaptive to the functional case. In the appendix, we prove that the asymptotic normality for the univariate BSD test statistic holds for the functional case. Firstly, we shall provide some mathematical preliminaries relevant to the asymptotic normality result. Then the BDS test statistic is related to a generalized $U$ statistic with order $2$. Finally, the asymptotic normality result and its proof are presented. It is worth noting that the following proof is not restricted by any distance measure when computing sup-norms.

\subsection{Mathematical Preliminaries}

The notation and definitions to be presented here follow those in \cite{Bosq99}. Let ${\cal H}$ be a separable Hilbert space with the inner product $\left < \cdot, \cdot \right >_H$ and the norm $|| \cdot ||_H$. We equip $\mathcal{H}$ with its Borel $\sigma$-field ${\cal B} (\mathcal{H})$. Since $\mathcal{H}$ is a linear metric space with a countable basis, it is a topological space with a countable basis for its topology. By Proposition 3.1 of \cite{Preston08}, since ${\cal B} (\mathcal{H})$ is the Borel $\sigma$-field of $\mathcal{H}$ generated by Borel subsets of $\mathcal{H}$, the measurable space $(\mathcal{H}, {\cal B} (\mathcal{H}))$ is countably generated. Let $(\Omega, {\cal F}, \mathbb{P})$ be a complete probability space. We consider a discrete-time functional time series $\{ X_t \}_{t \in {\cal Z}}$ on $(\Omega, {\cal F}, \mathbb{P})$ with values in the countably generated measurable space $(\mathcal{H}, {\cal B} (\mathcal{H}))$, where ${\cal Z}$ is the set of integers. Note that the condition that the state space of a stochastic process is a countably generated measurable space was imposed in \cite{DK83} in which some asymptotic normality results for $U$-statistics were presented. We shall use the asymptotic results for $U$ statistics to prove the asymptotic normality of the BDS test statistic here. Consequently, we also impose the condition that $\{ X_t \}_{t \in {\cal Z}}$ takes on values in the countably generated measurable space $(\mathcal{H}, {\cal B} (\mathcal{H}))$. As in the univariate case of \cite{broock1996test}, it is assumed that $\{ X_t \}_{t \in {\cal Z}}$ is a strictly stationary stochastic process. Let $\mu_{X_t}$ be a measure on $(\mathcal{H}, {\cal B} (\mathcal{H}))$ which is induced by the random element $X_t$, (i.e., an $\mathcal{H}$-valued random variable), under the measure $\mathbb{P}$. That is, for any $B \in {\cal B} (\mathcal{H})$, 
\begin{eqnarray} \label{imagemeasure}
\mu_{X_t} (B) := \mathbb{P} (X^{-1}_t (B)),
\end{eqnarray}
where $X^{-1}_t (B) := \{ \omega \in \Omega | X_t (\omega) \in B \} \in {\cal F}$. Note that $\mu_{X_t}$ is also called an image measure of $\mathbb{P}$ under the measurable map $X_t : \Omega \to \mathcal{H}$ \citep[see, e.g.,][]{PrakasaRao14}.

Under the assumption that $\{ X_t \}_{t \in {\cal Z}}$ a strictly stationary stochastic process, the image measure $\mu_{X_t}$ is time-invariant. Therefore, we write $\mu_{X}$ for $\mu_{X_t}$. Let $X^m_t := (X_t, X_{t+1}, \dots, X_{t+m-1}) \in \mathcal{H}^m$. Note that $X^m_t$ is a generalization of the $m$-history in \cite{broock1996test} from the univariate case to the functional case. When $\{ X_t \}_{t \in {\cal Z}}$ are independent under the measure $\mathbb{P}$, we consider the $m$-product countably generated measurable space $(\mathcal{H}^m, {\cal B} (\mathcal{H})^{\otimes m})$. In this case, the image measure $\mu_{X^m_t}$ of $\mathbb{P}$ under $X^m_t$ on $(\mathcal{H}^m, {\cal B} (\mathcal{H})^{\otimes m})$ is given by:
\begin{eqnarray} \label{imageproductmeasure}
\mu_{X^m_t} (B_1 \times B_2 \times \cdots \times B_m) = \prod^{m}_{i = 1} \mu_X (B_i), 
\end{eqnarray}
for any $B_1 \times B_2 \times \cdots \times B_m \in {\cal B} (\mathcal{H})^{\otimes m}$. Since $\mu_{X^m_t}$ is time invariant, we write $\mu_{X^m}$ for $\mu_{X^m_t}$.

For each $i, j = 1, 2, \cdots$ with $i < j$, let ${\cal G}_{i, j}$ be the $\mathbb{P}$-augmentation of the $\sigma$-field generated by the set of $\mathcal{H}$-valued random elements $\sigma (X_i, X_{i+1}, \cdots, X_j)$. Then, according to \cite{VR61}, \cite{grassberger1983characterization} and \cite{broock1996test}, the $\mathcal{H}$-valued stochastic process $\{ X_t \}_{t \in {\cal Z}}$ is said to be absolutely regular if 
\begin{eqnarray} \label{Regularbeta}
\beta (k) := \sup_{n \in \mathbb{N}} \bigg \{ \mbox{E} \bigg [ \sup \bigg \{ | \mathbb{P} (G | {\cal G}_{1, n} ) - \mathbb{P} (G) | \bigg | G \in {\cal G}_{n+k, \infty} \bigg \} \bigg ]  \bigg \},
\end{eqnarray}
converges to zero as $k \to \infty$, where $\mathbb{N}$ is the set of natural numbers.

For each ${\bf x} \in \mathcal{H}^m$, we consider the max-norm defined by: 
\begin{eqnarray} \label{max-norm}
|| {\bf x} ||_{m, \mathcal{H}} := \max_{k = 1, 2, \cdots, m} \{ || x_i ||_H  \},
\end{eqnarray} 
where $x_i \in \mathcal{H}$ and $|| \cdot ||_{\mathcal{H}}$ is the norm on $\mathcal{H}$. For the numerical implementation of the BDS test in the functional case, we consider the $L^2_H (\mathbb{P})$ space (i.e., the space of square-integrable $\mathcal{H}$-valued random elements under the measure $\mathbb{P}$). The notation $L^2_H (\mathbb{P})$ follows that in \cite{Bosq99}. Note that $L^2_H (\mathbb{P})$ is the space of $\mathcal{H}$-valued random elements on $(\Omega, {\cal F}, \mathbb{P})$ with the following norm:
\begin{eqnarray} \label{L2norm}
|| X ||_{L^2_H (\mathbb{P})} := \bigg ( \int_{\Omega} || X (\omega) ||^2_H \mathbb{P} (d \omega) \bigg )^{\frac{1}{2}} = \bigg ( \int_{\mathcal{H}} || x ||^2_H \mu_X (d x) \bigg )^{\frac{1}{2}}. 
\end{eqnarray}
In this case, the max-norm in~\eqref{max-norm} becomes:
\begin{eqnarray} \label{max-norm-L2}
|| {\bf x} ||_{m, L^2_H (\mathbb{P})} := \max_{k = 1, 2, \cdots, m} \{ || x_i ||_{L^2_H (\mathbb{P})}  \},
\end{eqnarray} 
Here we attempt to prove the asymptotic normality results for the BDS test statistic in the functional case for the general case of the max-norm in~\eqref{max-norm}.

Let $I_A$ be the characteristic function of a set $A$. In particular, when $A = [0, \epsilon)$, its characteristic function is, for simplicity, denoted by $I_{\epsilon}$, for any $\epsilon > 0$. We now extend the correlation integral in \cite{grassberger1983characterization} to the case of functional time series. Specifically, the correlation integral for the functional time series $\{ X_t \}_{t \in {\cal Z}}$ at embedding dimension $m$ is defined as:
\begin{eqnarray} \label{Cmn}
C_{m, n} (\epsilon) := \frac{1}{{n \choose 2}} \sum_{1 \le s \le t \le n} I_{\epsilon} (|| X^m_s - X^m_t ||_{m, \mathcal{H}}).  
\end{eqnarray}
As noted in \cite{broock1996test}, under the assumption that $\{ X_t \}_{t \in {\cal Z}}$ is a strictly stationary stochastic process that is regular, the limit expressed as $\lim_{n \to \infty} C_{m, n} (\epsilon)$ exists, and it is denoted by:
\begin{eqnarray} \label{Cm}
C_m (\epsilon) := \lim_{n \to \infty} C_{m, n} (\epsilon).
\end{eqnarray}
In the case of the functional time series, the limit in Eq. (\ref{Cm}) is given by:
\begin{eqnarray} \label{CmIntegrals}
C_m (\epsilon) = \int_{\mathcal{H}^m} \int_{\mathcal{H}^m} I_{\epsilon} (|| {\bf x} - {\bf y} ||_{m, \mathcal{H}}) \mu_{X^m} (d {\bf x}) \mu_{X^m} (d {\bf y}). 
\end{eqnarray}
When $\{ X_t \}_{t \in {\cal Z}}$ is an independent process, 
\begin{eqnarray} \label{Iproduct}
I_{\epsilon} (|| {\bf x} - {\bf y} ||_{m, \mathcal{H}}) = \prod^{m}_{i = 1} I_{\epsilon} (|| x_i - y_i  ||_{\mathcal{H}}).
\end{eqnarray}
Consequently, using~\eqref{imageproductmeasure} and~\eqref{Iproduct},~\eqref{CmIntegrals} becomes:
\begin{eqnarray} \label{CmIntegralsIndept}
C_m (\epsilon) &=& \int_{\mathcal{H}^m} \int_{\mathcal{H}^m} \bigg ( \prod^{m}_{i = 1} I_{\epsilon} (|| x_i - y_i  ||_{\mathcal{H}})  \bigg ) \prod^{m}_{i = 1} \mu_X (d x_i) \prod^{m}_{i = 1} \mu_X (d y_i) \nonumber\\
&=& \prod^{m}_{i = 1} \int_{\mathcal{H}} \int_{\mathcal{H}} I_{\epsilon} (|| x_i - y_i  ||_{\mathcal{H}}) \mu_X (d x_i) \mu_X (d y_i) = ( C_1 (\epsilon)  )^m. 
\end{eqnarray}
Write $C (\epsilon)$ for $C_1 (\epsilon)$. Then, from~\eqref{CmIntegralsIndept},
\begin{eqnarray} \label{CmIntegralsIndept1}
C_m (\epsilon) = [ C (\epsilon) ]^m. 
\end{eqnarray}

\subsection{Generalized $U$ Statistic with Order 2}

In the sequel, some concepts of a generalized $U$ statistic with order $2$ for the functional time series are presented. The notion of $U$-statistic may be dated back to \cite{Hoeffding48}. Here we extend the generalized $U$ statistic in \cite{Serfling1980}, and \cite{broock1996test} to the case of functional time series. The expositions here follow those in \cite{DK83} and \cite{broock1996test}.

Since $(\mathcal{H}, {\cal B} (\mathcal{H}))$ is a countably generated measurable space, the (finite) product space $(\mathcal{H}^m, {\cal B} (\mathcal{H})^{\otimes m})$ is also
a countably generated measurable space. Let $h : (\mathcal{H}^m)^2 \to \Re$ be a measurable function $h ({\bf x}, {\bf y})$, for ${\bf x}, {\bf y} \in \mathcal{H}^m$. The measurable function $h$ is called a kernel for the integral:
\begin{eqnarray} \label{kernelintegral}
\int_{\mathcal{H}^m} \int_{\mathcal{H}^m} h ({\bf x}, {\bf y}) \mu_{X^m} (d {\bf x}) \mu_{X^m} (d {\bf y}),
\end{eqnarray} 
if $h$ is symmetric in its arguments ${\bf x}$ and ${\bf y}$. That is,
\begin{eqnarray} \label{hsymmetric}
h ({\bf x}, {\bf y}) = h({\bf y}, {\bf x}). 
\end{eqnarray}
Then a generalized $U$-statistic with order $2$ is given by:
\begin{eqnarray} \label{GeneralizedUStatistic2}
U_n = \frac{1}{{n \choose 2}} \sum_{0 \le s \le t \le n} h(X^m_s, X^m_t), \quad n \ge 2.
\end{eqnarray}
Note that
\begin{eqnarray} \label{Isymmetric}
I_{\epsilon} (|| {\bf x} - {\bf y} ||_{m, \mathcal{H}}) = I_{\epsilon} (|| {\bf y} - {\bf x} ||_{m, \mathcal{H}}).
\end{eqnarray}
Then $I_{\epsilon} (|| {\bf x} - {\bf y} ||_{m, \mathcal{H}})$ is a symmetric kernel for the integral in~\eqref{CmIntegrals}. Consequently, by taking $h(X^m_s, X^m_t)$ in~\eqref{GeneralizedUStatistic2} as $I_{\epsilon} (||X^m_s  - X^m_t ||_{m, \mathcal{H}})$, the correlation integral in~\eqref{Cmn} coincides with~\eqref{GeneralizedUStatistic2}. Therefore, the correlation integral in~\eqref{Cmn} is a generalized $U$-statistic with order $2$ and symmetric kernel $I_{\epsilon} (|| {\bf x} - {\bf y} ||_{m, \mathcal{H}})$. Extending the definition in \cite{broock1996test} to the case of functional time series, we define:
\begin{eqnarray} \label{K}
K (\epsilon) &:=& \int_{\mathcal{H}} \bigg ( \int_{\mathcal{H}} I_{\epsilon} (|| x - y ||_H) \mu_X (d x) \bigg)^2 \mu_X (d y) \nonumber\\
&=& \int_{\mathcal{H}} \bigg ( \int_{\mathcal{H}} I_{\epsilon} (|| y - x ||_H) \mu_X (d y) \bigg)^2 \mu_X (d x).
\end{eqnarray}
The last equality follows symmetry. Then,
\begin{eqnarray} \label{VarEI}
&& \mbox{Var} [ \mbox{E} [ I_{\epsilon} (|| X_t - X_s ||_H ) | X_s ] ]  \nonumber\\
&=& \mbox{E} [ (\mbox{E} [ I_{\epsilon} (|| X_t - X_s ||_H) | X_s ] )^2 ] 
- ( \mbox{E} [\mbox{E} [I_{\epsilon} (|| X_t - X_s ||_H ) | X_s ]] )^2 \nonumber\\
&=& \mbox{E} [ (\mbox{E} [ I_{\epsilon} (|| X_t - X_s ||_H) | X_s ] )^2 ] 
- (\mbox{E} [I_{\epsilon} (|| X_t - X_s ||_H )] )^2 \nonumber\\
&=& \int_{\mathcal{H}} \bigg ( \int_{\mathcal{H}} I_{\epsilon} (|| x - y ||_H) \mu_X (d x) \bigg)^2 \mu_X (d y) 
- \bigg( \int_{\mathcal{H}} \int_{\mathcal{H}} I_{\epsilon} (|| x - y ||_H) \mu_X (d x) \mu_X (d y) \bigg )^2 \nonumber\\
&=& K(\epsilon) - [C(\epsilon)]^2.
\end{eqnarray}
Consequently, 
\begin{eqnarray}
K(\epsilon) \ge [C(\epsilon)]^2.
\end{eqnarray}

\subsection{Asymptotic Normality and its Proof}

To establish the asymptotic normality results for the generalized $U$ statistic with order $2$ in~\eqref{GeneralizedUStatistic2}, as in \cite{broock1996test}, we focus on the non-degenerate case where $K (\epsilon) > [C(\epsilon)]^2$. 

To simplify the notation, as in \cite{broock1996test}, we write $K$ for $K (\epsilon)$ and $C$ for $C (\epsilon)$ unless otherwise stated. Define $\sigma^2_m := \sigma^2_m (\epsilon)$ as follows:
\begin{eqnarray} \label{sigmam}
\sigma^2_m = 4 K^m - 4 C^{2 m} + 8 \sum^{m-1}_{i = 1} (K^{m-i} C^{2 i} - C^{2m}).
\end{eqnarray}

The following theorem gives the first asymptotic normality result, which extends \citet[][Theorem 2.1]{broock1996test} to the case of functional time series.

\begin{theorem} Suppose that
\begin{asparaenum}
\item $\{ X_t \}_{t \in {\cal Z}}$ is a sequence of IID $\mathcal{H}$-valued random elements;
\item $K (\epsilon) > [C(\epsilon)]^2$.
\end{asparaenum}
Then the standardized generalized $U$ statistic with order $2$ defined by:
\begin{eqnarray} \label{StandardizedU}
\sqrt{n} \bigg ( \frac{C_{m, n} (\epsilon) - (C (\epsilon))^m}{\sigma_m (\epsilon)} \bigg )
\end{eqnarray}
converges in distribution to $N(0, 1)$, (i.e., a standard normal distribution with zero mean and unit variance), as $n \to \infty$, where $\sigma_m (\epsilon)$ is given by~\eqref{sigmam}.
\end{theorem}

\begin{proof} 
The proof follows from Theorem 1(c) in \cite{DK83} and the proof of Theorem 2.1 in \cite{broock1996test}. Here we consider the $\mathcal{H}^m$-valued stochastic process $\{ X^m_t \}_{t \in {\cal Z}}$ on the probability space $(\Omega, {\cal F}, \mathbb{P})$. Under Condition 1 that $\{ X_t \}_{t \in {\cal T}}$ are IID, $X^m_s$ and $X^m_t$ are independent if $| s - t | \ge m$. Then the two $\sigma$-fields ${\cal G}_{1, n}$ and ${\cal G}_{n+k, \infty}$, for $k \ge m$, must be independent. Then for any $G \in {\cal G}_{n+k, \infty}$ with $k \ge m$,
\begin{eqnarray} \label{PG}
P (G | {\cal G}_{1, n}) = P (G).
\end{eqnarray}
Consequently, for each $k \ge m$, $\beta (k)$ in~\eqref{Regularbeta} must be identical to zero. This implies that $\{ X^m_t \}_{t \in {\cal Z}}$ is absolutely regular. Since it was assumed that $\{ X_t \}_{t \in {\cal Z}}$ is a strictly stationary $\mathcal{H}$-valued process, $\{ X^m_t \}_{t \in {\cal Z}}$ is an absolutely regular strictly stationary $\mathcal{H}$-valued process. Since $\beta (k) < \infty$, for all $k \le m$, and $\beta (k) = 0$, for all $k > m$, 
$\sum^{\infty}_{k = 1} [\beta (k)]^{\frac{\delta}{2 + \delta}} < \infty$, for some $\delta > 0$. Note that $I_{\epsilon} \le 1$. Then 
\begin{eqnarray} \label{CheckCondition}
\mbox{E} [ | I_{\epsilon} (|| X^m_s - X^m_t ||_{m, \mathcal{H}}) |^{2 + \delta} ] \le 1 < \infty.
\end{eqnarray}
Consequently, the conditions in \citet[][Theorem 1(c)]{DK83} are fulfilled. This then establishes the convergence of the standardized generalized $U$ statistic with order $2$ in~\eqref{StandardizedU} in distribution to a standard normal distribution. It remains to prove that $\sigma_m (\epsilon)$ is given by~\eqref{sigmam}. Define, for each ${\bf x} \in \mathcal{H}^m$, 
\begin{eqnarray} \label{h1}
h_1 ({\bf x}) &:=& \int_{\mathcal{H}^m} I_{\epsilon} (|| {\bf x} - {\bf y} ||_{m, \mathcal{H}}) \mu_{X^m} (d {\bf y}) - (C(\epsilon))^m \nonumber\\
&=& \prod^{m}_{i = 1} \int_{\mathcal{H}} I_{\epsilon} (|| x_i - y_i ||_{\mathcal{H}}) \mu_{X} (d y_i) - (C(\epsilon))^m. 
\end{eqnarray}
For each $i = 1, 2, \cdots, m$, let $X^m_t (i)$ be the $i^{th}$ component of $X^m_t$. Under Condition~1, 
\begin{eqnarray} \label{Expectedh1}
\mbox{E} [h_1 (X^m_t) ] &:=& \mbox{E} \bigg [ \prod^{m}_{i = 1} \int_{\mathcal{H}} I_{\epsilon} (|| X^m_t (i) - y_i ||_{\mathcal{H}}) \mu_{X} (d y_i) \bigg ] - (C(\epsilon))^m \nonumber\\
&=& \prod^{m}_{i = 1} \mbox{E} \bigg [ \int_{\mathcal{H}} I_{\epsilon} (|| X^m_t (i) - y_i ||_{\mathcal{H}}) \mu_{X} (d y_i) \bigg ] - (C(\epsilon))^m \nonumber\\
&=& \prod^{m}_{i = 1} \int_{\mathcal{H}} \int_{\mathcal{H}} I_{\epsilon} (|| x_i - y_i ||_{\mathcal{H}}) \mu_{X} (d x_i) \mu_{X} (d y_i) - (C(\epsilon))^m \nonumber\\
&=& \prod^{m}_{i = 1} C(\epsilon) - (C(\epsilon))^m = 0.  
\end{eqnarray}
Using the asymptotic variance from \cite{DK83} and writing
$\sigma^2_m$ for $\sigma^2_m (\epsilon)$,
\begin{eqnarray} \label{AsyVar}
\frac{1}{4} \sigma^2_m = \mbox{E} [(h_1 (X^m_1))^2] + 2 \sum_{t > 1} \mbox{E} [h_1 (X^m_1) h_1 (X^m_t)].
\end{eqnarray}
From~\eqref{h1},
\begin{eqnarray} \label{h1Sq}
&& (h_1 (X^m_t))^2  \nonumber\\
&=& \bigg ( \prod^{m}_{i = 1} \int_{\mathcal{H}} I_{\epsilon} (|| X^m_t (i) - y_i ||_{\mathcal{H}}) \mu_{X} (d y_i) - (C(\epsilon))^m \bigg )^2 \nonumber\\
&=& \bigg ( \prod^{m}_{i = 1} \int_{\mathcal{H}} I_{\epsilon} (|| X^m_t (i) - y_i ||_{\mathcal{H}}) \mu_{X} (d y_i) \bigg )^2 - 2 \bigg ( \prod^{m}_{i = 1} \int_{\mathcal{H}} I_{\epsilon} (|| X^m_t (i) - y_i ||_{\mathcal{H}}) \mu_{X} (d y_i) \bigg )  \nonumber\\
&& \times (C(\epsilon))^m + (C(\epsilon))^{2 m}.
\end{eqnarray} 
Taking expectation in~\eqref{h1Sq} gives:
\begin{eqnarray} \label{Expectedh1Sq}
\mbox{E} [ (h_1 (X^m_t))^2 ] &=& \mbox{E} \bigg [ \bigg ( \prod^{m}_{i = 1} \int_{\mathcal{H}} I_{\epsilon} (|| X^m_t (i) - y_i ||_{\mathcal{H}}) \mu_{X} (d y_i) \bigg )^2 \bigg ] \nonumber\\
&& - 2 \mbox{E} \bigg [ \bigg ( \prod^{m}_{i = 1} \int_{\mathcal{H}} I_{\epsilon} (|| X^m_t (i) - y_i ||_{\mathcal{H}}) \mu_{X} (d y_i) \bigg ) \bigg ] (C(\epsilon))^m \nonumber\\
&& + (C(\epsilon))^{2 m} \nonumber\\
&=& \int_{\mathcal{H}^m} \bigg ( \prod^{m}_{i = 1} \int_{\mathcal{H}} I_{\epsilon} (|| x_i - y_i ||_{\mathcal{H}}) \mu_{X} (d y_i) \bigg )^2 \mu_{X^m} (d {\bf x} ) \nonumber\\
&& - 2 \prod^{m}_{i = 1} \mbox{E} \bigg [ \bigg (  \int_{\mathcal{H}} I_{\epsilon} (|| X^m_t (i) - y_i ||_{\mathcal{H}}) \mu_{X} (d y_i) \bigg ) \bigg ] (C(\epsilon))^m \nonumber\\
&& + (C(\epsilon))^{2 m} \nonumber\\
&=& \prod^{m}_{i = 1}  \int_{\mathcal{H}} \bigg ( \int_{\mathcal{H}} I_{\epsilon} (|| x_i - y_i ||_{\mathcal{H}}) \mu_{X} (d y_i) \bigg )^2 \mu_{X} (d x_i ) \nonumber\\
&& - 2 \prod^{m}_{i = 1} \mbox{E} \bigg [ \bigg (  \int_{\mathcal{H}} I_{\epsilon} (|| X^m_t (i) - y_i ||_{\mathcal{H}}) \mu_{X} (d y_i) \bigg ) \bigg ] (C(\epsilon))^m \nonumber\\
&& + (C(\epsilon))^{2 m} \nonumber\\
&=& \prod^{m}_{i = 1} K (\epsilon) - 2 \prod^{m}_{i = 1} C (\epsilon) (C(\epsilon))^m + (C(\epsilon))^{2 m} \nonumber\\
&=& (K (\epsilon))^m - (C(\epsilon))^{2 m}. 
\end{eqnarray}
To evaluate $\mbox{E} [h_1 (X^m_1) h_1 (X^m_t)]$, two cases are considered, namely $t \le m$ and $t > m$. 

\noindent
Recall that $X^m_1 = (X_1, X_2, \cdots, X_m) \in \mathcal{H}^m$ and $X^m_t = (X_t, X_{t+1}, \cdots, X_{t+m-1}) \in \mathcal{H}^m$. Then for $t \le m$, the overlapping elements of $X^m_1$ and $X^m_t$ are $(X_t, X_{t+1}, \cdots, X_{m}) \in \mathcal{H}^{m-t+1}$. The non-overlapping elements are:
\begin{eqnarray}
&& (X_1, X_2, \cdots, X_{t-1}) \in \mathcal{H}^{t-1}, \nonumber\\
&& (X_{m+1}, X_{m+2}, \cdots, X_{t+m-1}) \in \mathcal{H}^{t-1}.
\end{eqnarray}
Consequently, for $t \le m$, 
\begin{eqnarray} \label{Expectedh1product}
&& \mbox{E} [h_1 (X^m_1) h_1 (X^m_t)] \nonumber\\
&=& \mbox{E} \bigg [ \bigg ( \prod^{m}_{i = 1} \int_{\mathcal{H}} I_{\epsilon} (|| X_i - y_i ||_{\mathcal{H}}) \mu_{X} (d y_i) - (C(\epsilon))^m \bigg ) \nonumber\\
&& \times \bigg ( \prod^{t+m-1}_{i = t} \int_{\mathcal{H}} I_{\epsilon} (|| X_i - y_i ||_{\mathcal{H}}) \mu_{X} (d y_i) - (C(\epsilon))^m  \bigg ) \bigg ] \nonumber\\
&=& \mbox{E} \bigg [ \bigg ( \prod^{m}_{i = 1} \int_{\mathcal{H}} I_{\epsilon} (|| X_i - y_i ||_{\mathcal{H}}) \mu_{X} (d y_i) \bigg ) \bigg ( \prod^{t+m-1}_{i = t} \int_{\mathcal{H}} I_{\epsilon} (|| X_i - y_i ||_{\mathcal{H}}) \mu_{X} (d y_i)  \bigg ) \bigg ] \nonumber\\
&& - (C(\epsilon))^m  \mbox{E} \bigg [ \bigg ( \prod^{m}_{i = 1} \int_{\mathcal{H}} I_{\epsilon} (|| X_i - y_i ||_{\mathcal{H}}) \mu_{X} (d y_i) \bigg ) \bigg] \nonumber\\
&& - (C(\epsilon))^m \mbox{E} \bigg [ \bigg ( \prod^{t+m-1}_{i = t} \int_{\mathcal{H}} I_{\epsilon} (|| X_i - y_i ||_{\mathcal{H}}) \mu_{X} (d y_i) \bigg ) \bigg ] + (C(\epsilon))^{2m} \nonumber\\
&=& \mbox{E} \bigg [ \bigg ( \prod^{t-1}_{i = 1} \int_{\mathcal{H}} I_{\epsilon} (|| X_i - y_i ||_{\mathcal{H}}) \mu_{X} (d y_i) \bigg ) \bigg ( \prod^{m}_{i = t} \int_{\mathcal{H}} I_{\epsilon} (|| X_i - y_i ||_{\mathcal{H}}) \mu_{X} (d y_i) \bigg )^2 \nonumber\\
&& \times \bigg ( \prod^{t+m-1}_{i = m+1} \int_{\mathcal{H}} I_{\epsilon} (|| X_i - y_i ||_{\mathcal{H}}) \mu_{X} (d y_i)  \bigg ) \bigg ] \nonumber\\
&& - (C(\epsilon))^m  \prod^{m}_{i = 1} \mbox{E} \bigg [ \int_{\mathcal{H}} I_{\epsilon} (|| X_i - y_i ||_{\mathcal{H}}) \mu_{X} (d y_i) \bigg] \nonumber\\
&& - (C(\epsilon))^m \prod^{t+m-1}_{i = t} \mbox{E} \bigg [ \int_{\mathcal{H}} I_{\epsilon} (|| X_i - y_i ||_{\mathcal{H}}) \mu_{X} (d y_i) \bigg ] + (C(\epsilon))^{2m} \nonumber\\
\end{eqnarray}
Note that
\begin{eqnarray} \label{CmExp}
&& \mbox{E} \bigg [ \int_{\mathcal{H}} I_{\epsilon} (|| X_i - y_i ||_{\mathcal{H}}) \mu_{X} (d y_i) \bigg] \nonumber\\
&=& \int_{\mathcal{H}} \int_{\mathcal{H}} I_{\epsilon} (|| x_i - y_i ||_{\mathcal{H}}) \mu_{X} (d y_i) \mu_{X} (d x_i) = C(\epsilon), 
\end{eqnarray}
and that
\begin{eqnarray} \label{Expectedh1productFirstTerm}
&& \mbox{E} \bigg [ \bigg ( \prod^{t-1}_{i = 1} \int_{\mathcal{H}} I_{\epsilon} (|| X_i - y_i ||_{\mathcal{H}}) \mu_{X} (d y_i) \bigg ) \bigg ( \prod^{m}_{i = t} \int_{\mathcal{H}} I_{\epsilon} (|| X_i - y_i ||_{\mathcal{H}}) \mu_{X} (d y_i) \bigg )^2 \nonumber\\
&& \times \bigg ( \prod^{t+m-1}_{i = m+1} \int_{\mathcal{H}} I_{\epsilon} (|| X_i - y_i ||_{\mathcal{H}}) \mu_{X} (d y_i)  \bigg ) \bigg ] \nonumber\\
&=& \int_{\mathcal{H}^{t+m-1}} \bigg ( \prod^{t-1}_{i = 1} \int_{\mathcal{H}} I_{\epsilon} (|| x_i - y_i ||_{\mathcal{H}}) \mu_{X} (d y_i) \bigg ) \prod^{m}_{i = t}  \bigg (\int_{\mathcal{H}} I_{\epsilon} (|| x_i - y_i ||_{\mathcal{H}}) \mu_{X} (d y_i) \bigg )^2 \nonumber\\
&& \times \bigg ( \prod^{t+m-1}_{i = m+1} \int_{\mathcal{H}} I_{\epsilon} (|| x_i - y_i ||_{\mathcal{H}}) \mu_{X} (d y_i)  \bigg ) \mu_{X^{t+m-1}} (d {\bf x}) \nonumber\\
&=& \bigg ( \prod^{t-1}_{i = 1} \int_{\mathcal{H}} \int_{H} I_{\epsilon} (|| x_i - y_i ||_{\mathcal{H}}) \mu_{X} (d y_i) \mu_{X} (d x_i)  \bigg ) \nonumber\\
&& \times \prod^{m}_{i = t} \int_{\mathcal{H}} \bigg ( \int_{\mathcal{H}} I_{\epsilon} (|| x_i - y_i ||_{\mathcal{H}}) \mu_{X} (d y_i) \bigg )^2 \mu_{X} (d x_i) \nonumber\\
&& \times \bigg ( \prod^{t+m-1}_{i = m+1} \int_{\mathcal{H}} \int_{\mathcal{H}} I_{\epsilon} (|| x_i - y_i ||_{\mathcal{H}}) \mu_{X} (d y_i) \mu_{X} (d x_i)  \bigg ) \nonumber\\
&=& \bigg ( \prod^{t-1}_{i = 1} C (\epsilon) \bigg ) \bigg ( \prod^{m}_{i = t} K(\epsilon) 
\bigg ) \bigg ( \prod^{t+m-1}_{i = m+1}  C (\epsilon) \bigg ) \nonumber\\
&=& (C(\epsilon))^{t-1} (K(\epsilon))^{m-t+1} (C(\epsilon))^{t-1} \nonumber\\
&=& (K(\epsilon))^{m-(t-1)} (C(\epsilon))^{2(t-1)}. 
\end{eqnarray}
Therefore, using Eq. (\ref{Expectedh1product}), Eq. (\ref{CmExp}) and Eq. (\ref{Expectedh1productFirstTerm}), for $t \le m$, 
\begin{eqnarray} \label{Expectedh1product1}
\mbox{E} [h_1 (X^m_1) h_1 (X^m_t)] = (K(\epsilon))^{m-(t-1)} (C(\epsilon))^{2(t-1)} 
- (C(\epsilon))^{2m}.
\end{eqnarray}
For $t > m$, $X^m_1$ and $X^m_t$ are independent because they do not have overlapping terms. Using this fact and~\eqref{Expectedh1}, 
\begin{eqnarray} \label{Expectedh1product0} 
\mbox{E} [h_1 (X^m_1) h_1 (X^m_t)] = \mbox{E} [h_1 (X^m_1) ] \mbox{E} [  h_1 (X^m_t)]  = 0.
\end{eqnarray}
Consequently, using~\eqref{AsyVar},~\eqref{Expectedh1Sq},~\eqref{Expectedh1product1} and~\eqref{Expectedh1product0},
\begin{eqnarray} \label{AsyVar1}
\frac{1}{4} \sigma^2_m &=& \mbox{E} [(h_1 (X^m_1))^2] + 2 \sum_{t > 1} \mbox{E} [h_1 (X^m_1) h_1 (X^m_t)] \nonumber\\
&=& \mbox{E} [(h_1 (X^m_1))^2] + 2 \sum^{\infty}_{t = 2} \mbox{E} [h_1 (X^m_1) h_1 (X^m_t)] \nonumber\\
&=& \mbox{E} [(h_1 (X^m_1))^2] + 2 \sum^{m}_{t = 2} \mbox{E} [h_1 (X^m_1) h_1 (X^m_t)] \nonumber\\
&=& (K (\epsilon))^m - (C(\epsilon))^{2 m} + 2 \sum^{m}_{t = 2} \bigg ( (K(\epsilon))^{m-(t-1)} (C(\epsilon))^{2(t-1)} - (C(\epsilon))^{2m} \bigg ) \nonumber\\
&=& (K (\epsilon))^m - (C(\epsilon))^{2 m} + 2 \sum^{m-1}_{t = 1} \bigg ( (K(\epsilon))^{m-t} (C(\epsilon))^{2t} - (C(\epsilon))^{2m} \bigg ) \nonumber\\
&=& K^m - C^{2 m} + 2 \sum^{m-1}_{t = 1} ( K^{m-t} C^{2t} - C^{2m}) 
\end{eqnarray}
This then gives the asymptotic variance in~\eqref{sigmam} and completes the proof.
\end{proof}
\newpage
\bibliographystyle{agsm}
\bibliography{reference.bib}

@Article{packard1980geometry,
  author    = {Packard, Norman H and Crutchfield, James P and Farmer, J Doyne and Shaw, Robert S},
  journal   = {Physical Review Letters},
  title     = {Geometry from a time series},
  year      = {1980},
  number    = {9},
  pages     = {712},
  volume    = {45},
  publisher = {APS},
}

@Article{HS23,
  author = {X. Huang and H. L. Shang},
  journal = {Nonlinear Dynamics},
  title = {Nonlinear autocorrelation function of functional time series},
  year = {2023},
  volume = {111},
  pages = {2537-2554}
}

@InCollection{takens1981detecting,
  author    = {Takens, Floris},
  booktitle = {Dynamical Systems and Turbulence, Warwick 1980},
  publisher = {Springer},
  title     = {Detecting strange attractors in turbulence},
  year      = {1981},
  pages     = {366--381},
}

@Article{grassberger1983characterization,
  author    = {Grassberger, Peter and Procaccia, Itamar},
  journal   = {Physical Review Letters},
  title     = {Characterization of strange attractors},
  year      = {1983},
  number    = {5},
  pages     = {346},
  volume    = {50},
  publisher = {APS},
}

@Book{brock1991nonlinear,
  author    = {Brock, William A and Hsieh, David Arthur and LeBaron, Blake Dean and Brock, William E and others},
  publisher = {MIT Press},
  title     = {{Nonlinear Dynamics, Chaos, and Instability: Statistical Theory and Economic Evidence}},
  year      = {1991},
  address   = {Cambridge, Massachusetts, London, England},
}

@Article{brock1987notes,
  author  = {Brock, W},
  journal = {Unpublished manuscript. Madison: University of Wisconsin},
  title   = {Notes on nuisance parameter problems in BDS type tests for IID},
  year    = {1987},
}

@Article{broock1996test,
  author    = {Broock, William A and Scheinkman, Jos{\'e} Alexandre and Dechert, W Davis and LeBaron, Blake},
  journal   = {Econometric Reviews},
  title     = {A test for independence based on the correlation dimension},
  year      = {1996},
  number    = {3},
  pages     = {197--235},
  volume    = {15},
  publisher = {Taylor \& Francis},
}

@Article{KST08,
  author = {E. Konstantinidi and G. Skiadopoulos and E. Tzagkaraki},
  journal = {Journal of Banking \& Finance},
  title = {{Can the evolution of implied volatility be forecasted? Evidence from European and US implied volatility indices}},
  year = {2008},
  volume = {32},
  number = {11},
  pages = {2401-2411}
}

@Article{FMS14,
  author = {M. Fernandes and M. C. Medeiros and M. Scharth},
  journal = {Journal of Banking \& Finance},
  title = {Modeling and predicting the {CBOE} market volatility index},
  year = {2014},
  volume = {40},
  pages = {1-10}
}

@Article{SK19,
  author = {H. L. Shang and Y. Yang and F. Kearney},
  title = {Intraday forecasts of a volatility index: Functional time series methods with dynamic updating},
  journal = {Annals of Operations Research},
  year = {2019},
  volume = {282},
  number = {1},
  pages = {331-354}
}

@Manual{serge22,
   title = {far: Modelization for Functional AutoRegressive Processes},
   author = {Damon Julien Guillas Serge},
   year = {2022},
   note = {R package version 0.6-6},
   url = {\url{https://CRAN.R-project.org/package=far}}
 }

@Article{f1996nuisance,
  author    = {Pedro J.F. de Lima},
  journal   = {Econometric Reviews},
  title     = {Nuisance parameter free properties of correlation integral based statistics},
  year      = {1996},
  number    = {3},
  pages     = {237--259},
  volume    = {15},
  publisher = {Taylor \& Francis},
}

@Article{aue2017functional,
  author    = {Aue, Alexander and Horv{\'a}th, Lajos and Pellatt, D. F.},
  journal   = {Journal of Time Series Analysis},
  title     = {Functional generalized autoregressive conditional heteroskedasticity},
  year      = {2017},
  number    = {1},
  pages     = {3--21},
  volume    = {38},
  publisher = {Wiley Online Library},
}

@Article{Bosq99,
  author  = {D. Bosq},
  journal = {Annales de l'ISUP, Publications de l'Institut de Statistique de l'Universit\'{e} de Paris},
  title   = {{Autoregressive Hilbertian processes}},
  year    = {1999},
  number  = {2-3},
  pages   = {25-55},
  volume  = {XXXXIII},
}

@TechReport{RWZ21,
  author      = {G. Rice and T. Wirjanto and Y. Zhao},
  institution = {University of Waterloo},
  title       = {{Exploring volatility of crude oil intra-day return curves: A functional GARCH-X model}},
  year        = {2021},
  number      = {109231},
  type        = {MPRA working paper},
  url = {\url{https://mpra.ub.uni-muenchen.de/109231/}}
}

@Manual{Team22,
  title = {R: A Language and Environment for Statistical Computing},
  author = {{R Core Team}},
  organization = {R Foundation for Statistical Computing},
  address = {Vienna, Austria},
  year = {2023},
  url = {\url{https://www.R-project.org/}}
}

@Article{SHX22,
  author  = {H. L. Shang and S. Haberman and R. Xu},
  journal = {Insurance: Mathematics and Economics},
  title   = {Multi-population modelling and forecasting life-table death counts},
  year    = {2022},
  pages   = {239-253},
  volume  = {106},
}

@Article{SCS22,
  author  = {H. L. Shang and J. Cao and P. Sang},
  journal = {Journal of the Royal Statistical Society: Series C},
  title   = {{Stopping time detection of wood panel compression: A functional time-series approach}},
  year    = {2022},
  volume  = {71},
  number = {5},
  pages = {1205-1224}
}

@Article{Hooker2020,
  author    = {Giles Hooker and Hanlin Shang},
  journal   = {Statistics and Computing},
  title     = {Selecting the derivative of a functional covariate in scalar-on-function regression},
  year      = {2022},
  number    = {3},
  pages     = {Article: 35},
  volume    = {32},
  publisher = {Springer, Springer Nature},
}

@Article{shang2019visualizing,
  author    = {Shang, Han Lin},
  journal   = {Journal of the Royal Statistical Society: Series A (Statistics in Society)},
  title     = {Visualizing rate of change: {An} application to age-specific fertility rates},
  year      = {2019},
  number    = {1},
  pages     = {249--262},
  volume    = {182},
  publisher = {Wiley Online Library},
}

@Book{kokoszka2017introduction,
  author    = {Kokoszka, Piotr and Reimherr, Matthew},
  publisher = {Chapman and Hall/CRC},
  title     = {Introduction to Functional Data Analysis},
  year      = {2017},
  address   = {Boca Raton},
}

@Book{ramsey2005functional,
  author    = {Ramsay, Jim O and Silverman, Bernard W},
  publisher = {Springer},
  title     = {Functional Data Analysis},
  year      = {2005},
  address   = {New York},
  journal   = {Springer Series in Statistics, New York: Springer Verlag},
}

@Article{kokoszka2017inference,
  author    = {Kokoszka, Piotr and Rice, Gregory and Shang, Han Lin},
  journal   = {Journal of Multivariate Analysis},
  title     = {Inference for the autocovariance of a functional time series under conditional heteroscedasticity},
  year      = {2017},
  pages     = {32--50},
  volume    = {162},
  publisher = {Elsevier},
}

@Book{ramsay2002applied,
  author    = {Ramsay, James O and Silverman, Bernard W},
  publisher = {Springer},
  title     = {Applied Functional Data Analysis: Methods and Case Studies},
  year      = {2002},
  address   = {New York},
  volume    = {77},
}

@Article{damon2002inclusion,
  author    = {Damon, Julien and Guillas, Serge},
  journal   = {Environmetrics},
  title     = {The inclusion of exogenous variables in functional autoregressive ozone forecasting},
  year      = {2002},
  number    = {7},
  pages     = {759--774},
  volume    = {13},
  publisher = {Wiley Online Library},
}

@InCollection{turbillon2007estimation,
  author    = {Turbillon, C{\'e}line and Marion, Jean-Marie and Pumo, Besnik},
  booktitle = {Recent advances in stochastic modeling and data analysis},
  publisher = {World Scientific},
  title     = {Estimation of the moving-average operator in a {H}ilbert space},
  year      = {2007},
  editor = {C. H. Skiadas},
  pages     = {597--604},
}

@Article{klepsch2017prediction,
  author    = {Klepsch, Johannes and Kl{\"u}ppelberg, Claudia and Wei, Taoran},
  journal   = {Econometrics and Statistics},
  title     = {Prediction of functional {ARMA} processes with an application to traffic data},
  year      = {2017},
  pages     = {128--149},
  volume    = {1},
  publisher = {Elsevier},
}

@article{zamani2022seasonal,
  title={Seasonal functional autoregressive models},
  author={Zamani, Atefeh and Haghbin, Hossein and Hashemi, Maryam and Hyndman, Rob J},
  journal={Journal of Time Series Analysis},
  volume={43},
  number={2},
  pages={197--218},
  year={2022},
  publisher={Wiley Online Library},
}

@Article{gonzalez2017forecasting,
  author    = {Gonz{\'a}lez, Jos{\'e} Portela and San Roque, Antonio Mu{\~n}oz San Mu{\~n}oz and Perez, Estrella Alonso},
  journal   = {IEEE Transactions on Power Systems},
  title     = {Forecasting functional time series with a new {H}ilbertian {ARMAX} model: Application to electricity price forecasting},
  year      = {2017},
  number    = {1},
  pages     = {545--556},
  volume    = {33},
  publisher = {IEEE},
}

@Article{hormann2013functional,
  author    = {H{\"o}rmann, Siegfried and Horv{\'a}th, Lajos and Reeder, Ron},
  journal   = {Econometric Theory},
  title     = {A functional version of the {ARCH} model},
  year      = {2013},
  number    = {2},
  pages     = {267--288},
  volume    = {29},
  publisher = {Cambridge University Press},
}

@Article{horvath2013test,
  author    = {Horv{\'a}th, Lajos and Hu{\v{s}}kov{\'a}, Marie and Rice, Gregory},
  journal   = {Journal of Multivariate Analysis},
  title     = {Test of independence for functional data},
  year      = {2013},
  pages     = {100--119},
  volume    = {117},
  publisher = {Elsevier},
}

@Article{zhang2016white,
  author    = {Zhang, Xianyang},
  journal   = {Journal of Econometrics},
  title     = {White noise testing and model diagnostic checking for functional time series},
  year      = {2016},
  number    = {1},
  pages     = {76--95},
  volume    = {194},
  publisher = {Elsevier},
}

@Article{mestre2021functional,
  author    = {Mestre, Guillermo and Portela, Jos{\'e} and Rice, Gregory and San Roque, Antonio Mu{\~n}oz and Alonso, Estrella},
  journal   = {Computational Statistics \& Data Analysis},
  title     = {Functional time series model identification and diagnosis by means of auto-and partial autocorrelation analysis},
  year      = {2021},
  pages     = {107108},
  volume    = {155},
  publisher = {Elsevier},
}

@incollection{bosq1991modelization,
  title={Modelization, nonparametric estimation and prediction for continuous time processes},
  author={Bosq, Denis},
  booktitle={Nonparametric functional estimation and related topics},
  pages={509--529},
  year={1991},
  publisher={Springer}
}

@article{lebaron1997fast,
  title={A fast algorithm for the {BDS} statistic},
  author={LeBaron, Blake},
  journal={Studies in Nonlinear Dynamics \& Econometrics},
  volume={2},
  number={2},
  year={1997},
  publisher={De Gruyter}
}

@Article{DK83,
  author  = {M. Denker and G. Keller},
  journal = {{Zeitschrift f\"{u}r Wahrscheinlichkeitstheorie und Verwandte Gebiete}},
  title   = {{On $U$-statistics and v. mises' statistics for weakly dependent processes}},
  year    = {1983},
  pages   = {505-522},
  volume  = {64},
}

@Article{Hoeffding48,
  author  = {W. Hoeffding},
  journal = {Annals of Mathematical Statistics},
  title   = {A class of statistics with asymptotically normal distribution},
  year    = {1948},
  number  = {3},
  pages   = {293-325},
  volume  = {19},
}

@Article{PrakasaRao14,
  author  = {B. L. S. {Prakasa Rao}},
  journal = {Journal of Multivariate Analysis},
  title   = {{Characterization of Gaussian distribution on a Hilbert space from samples of random size}},
  year    = {2014},
  pages   = {209-214},
  volume  = {132},
}

@TechReport{Preston08,
  author      = {C. Preston},
  institution = {arXiv},
  title       = {Some notes on standard {Borel} and related spaces},
  year        = {2008},
  type        = {Working paper},
  url         = {\url{https://arxiv.org/pdf/0809.3066.pdf}},
}

@Book{Serfling1980,
  author    = {R. J. Serfling},
  publisher = {John Wiley \& Sons},
  title     = {Approximation Theorems of Mathematical Statistics},
  year      = {1980},
  address   = {New York},
}

@Article{VR61,
  author  = {V. A. Volkonski and Y. A. Rozanov},
  journal = {Theory of Probability and its Applications},
  title   = {{Some limit theorems for random function II.}},
  year    = {1961},
  pages   = {186-198},
  volume  = {6},
}

@Book{bosq2000linear,
  author    = {Bosq, Denis},
  publisher = {Springer Science \& Business Media},
  title     = {Linear Processes in Function Spaces: Theory and Applications},
  year      = {2000},
  address   = {New York},
  volume    = {149},
}

@article{gabrys2007portmanteau,
  title={Portmanteau test of independence for functional observations},
  author={Gabrys, Robertas and Kokoszka, Piotr},
  journal={Journal of the American Statistical Association: Theory and Methods},
  volume={102},
  number={480},
  pages={1338--1348},
  year={2007},
  publisher={Taylor \& Francis}
}

@Article{fernandes2012finite,
  author  = {Fernandes, Marcelo and Preumont, Pierre-Yves},
  journal = {Brazilian Review of Econometrics},
  title   = {The finite-sample size of the {BDS} test for {GARCH} standardized residuals},
  year    = {2012},
  number  = {2},
  pages   = {241--260},
  volume  = {32},
}

\end{document}